\documentclass[10pt]{elsarticle}
\usepackage[center]{qtree}
\usepackage{lineno,hyperref}
\usepackage{amsmath}
\usepackage{amssymb}
\usepackage{float}
\usepackage{latexsym}
\usepackage{amsmath,amsthm}
\usepackage{amsfonts}
\usepackage{amssymb}
\usepackage{graphicx}
\usepackage{color}

 \journal{Communications in Nonlinear Science and Numerical Simulation}

\modulolinenumbers[1]

\usepackage{amsthm}

\begin{document}
\begin{frontmatter}
\title{Autonomous choices  among  deterministic evolution--laws as source of uncertainty}
\author[1]{Leonardo Trujillo\footnote{Corresponding author e-mail: leonardo.trujillo@gmail.com}}
\author[2]{Arnaud~Meyroneinc}
\author[1]{Kilver Campos}
\author[1]{Otto Rend\'on}
\author[3]{Leonardo Di G. Sigalotti}
\address[1]{Centro de F{\'{i}}sica, Instituto Venezolano de Investigaciones Cient{\'{i}}ficas, (IVIC), A.P. 20632, Caracas 1020--A, Venezuela}
\address[2]{Departamento de Matem\'aticas, Instituto Venezolano de Investigaciones Cient{\'{i}}ficas, (IVIC), A.P. 20632, Caracas 1020--A, Venezuela}
\address[3]{\'Area de F\'isica de Procesos Irreversibles, Departamento de Ciencias B\'asicas,  Universidad Aut\'onoma Metropolitana - Azcapotzalco (UAM-A), Av. San Pablo, 180, 02200 Mexico City, Mexico}

\begin{abstract}
We provide evidence of an extreme form of sensitivity to initial conditions in a family of one-dimensional self-ruling dynamical systems.
We prove that some hyperchaotic sequences are closed-form expressions of the orbits of these pseudo-random dynamical systems.
Each chaotic system in this family exhibits a sensitivity to initial conditions that encompasses the sequence of choices of the evolution rule in some collection of maps.
This opens a possibility to extend current theories of complex behaviors on the basis of intrinsic uncertainty in deterministic chaos.
\end{abstract}
\end{frontmatter}

\section{Introduction}
Deterministic chaos is founded on the mathematical formalism provided by the theory of dynamical systems, where time may be discrete or continuous~\cite{R98}.
Numerous processes in nature, engineering~\cite{E07}, economy~\cite{D83}, and social sciences~\cite{DHD13} are well described by discrete time dynamical systems. 
In spite of their deterministic character, many nonlinear discrete time dynamical systems can exhibit complicated behaviors,  
for which predictability is limited in a rather fundamental way.
The sensitivity to initial conditions, one of the prevailing paradigms in chaos theory, states that initially nearby trajectories can lead to very different sequences of states after a certain time~\cite{R98,E07}. 
The purpose of this paper is to provide numerical evidence of an extreme form of sensitivity  to initial conditions  in a family of self--ruling one-dimensional dynamical systems. 
This occurs through the autonomous choices of the evolution rule that are partially  encoded in the initial conditions.
Thereby, we also provide a mathematical insight into an extended family of systems beyond the canonical examples of chaotic Chebyshev maps, and into the correspondence between the structure of computer generated orbits and exact analytic orbits.

Demonstration of  the existence of such kind of self-ruling dynamical systems has practical and theoretical implications.
Theoretically, it could open a possibility to extend theories of complex behaviors and go beyond the known source of indeterminacy at the edge of deterministic chaos.
For instance, 
such systems have been observed in terms of ``return time'' maps from experimental data of laser intensities, and in their models, where their
mixed mode oscillations  phenomenology have been interpreted as the manifestation  of homoclinic chaos \cite{PFA91,ZBCAGA98}.
However, the current state of the theory for this kind of maps is not sufficiently developed to explain such dynamics in a wider class of systems.
In this direction, the present work represents a first step toward more general geometrical models.
Also, on a practical level, it should provide better models of pseudo-indeterminacy for the computational emulation of randomness~\cite{SF98}.
The dynamical system presented in this work 
may be used to enforce cryptosystem  security with potential applications in chaos-based  cryptography~\cite{K11}.

The paper is organized as follows.
In Section 2, we recall well-known concepts about chaotic dynamics that will be revisited in the following of the paper.
We also recall some aspects of the discrete family of Chebyshev polynomial maps, which have been studied extensively in the literature.
In Section 3, we embed these particular maps in a continuous one-parameter family of systems and provide a dynamical framework for their numerical and analytical study. We start with rational values of the parameter, which already allows us to correct some misinterpretations that appeared in the previous literature.
In Section 4 we further extend our numerical investigation and consider irrational values of the parameter.
Finally in Section 5, we discuss some conceptual implications of this work.

\section{Preliminaries}\label{Prem}

\subsection{Dynamical systems, determinism and chaos}

For many  processes, 
the  time evolution of a system can be described by values $x_n=x(t_n)$ of its dynamical state variables at a discrete sequence of times
$t_n$ ($n=0,1,2,\ldots$). 
A discrete time dynamical system intends to mathematically reproduce such sequences and is traditionally formalised as the repeated composition of a function (a map or transformation) $f:X \rightarrow X$ with itself ($f^{n+1}:=f\circ f^{n}$) over the state space~\cite{R98}. 
We recall that a function is a {\it rule} which assigns to each element $x$ a {\it unique} element $f(x)$.
Hence, if time is viewed as a discrete parameter, once the initial value $x_0$ is specified, the iterated composition of functions can be expressed as a first--order recurrence equation or map:
$x_{n+1}=f(x_n)$, $x_0\in X$, implying that  $x_n=f^{n}(x_0)$ for all $n\geq 0$.
This time evolution--law determines the state of the system at each instant $n$ from its state at a previous time. 
In this case, a forward sequence  defined by the set 
$\left\{ f^{n}(x_0)\right\}_{n=0}^{\infty}=\{ x_0,x_1,x_2,\dots \}=\mathcal{O}(x_0)$
is named the {\it orbit} and is uniquely determined by the initial condition $x_0$~\cite{R98}.

The very simple concept of iterated function (and the continuous time equivalent given by differential equations) allows the quantitative statement of the Newtonian determinism~\cite{Arnold89}.
The contributions of H. Poincar\'e at the end of the nineteenth century  marked a turning point in the understanding of this determinism, 
opening a new field to investigate the limits of predictability in many macroscopic phenomena.
If there exists a limit on the knowledge of the initial condition, it could happen that due to a small discrepancy  $\delta>0$ between  $x_0$ and $y_0$, the orbits
$\mathcal{O}(x_0)$ and $\mathcal{O}(y_0)$ could significantly differ from each other after some instant $N(\delta)<\infty$.
Because of this sensitive dependence on initial conditions, some deterministic and simple mathematical models may appear to behave non--deterministically.
However, once an initial condition is given, the iterative process determines a unique orbit~\cite{R98}. 
In what follows (starting in Section 3), we consider a family of one-dimensional systems where in an apparent paradox, many different orbits seem to originate from an initial condition.

\subsection{Chaotic maps with orbits having closed--form expressions}
In very few instances, chaotic dynamical systems have orbits that can be expressed 
in  closed--form, i.e., in terms of elementary algebraic and transcendental functions.
In the next subsections we will review such a well-known discrete family of maps, which will be relevant in what follows.

\subsubsection{The canonical Ulam-von Neumann map}
In 1947, S.M. Ulam and J. von Neumann~\cite{UvN47}
understood  that by iterating the function 
$f(x)=4x(1-x)$, $x\in [0,1]$, it is possible to produce  real-valued sequences with very
complex patterns, emulating a random process. In the literature the quadratic transformations $x\mapsto 4x(1-x)$, $x\in[0,1]$ and $x\mapsto 1-2x^2$, $x\in[-1,1]$, are usually referred  to as Ulam--von Neumann maps, see  Refs.~\cite{U97} and \cite{J95}.
A 1976 article by R. May~\cite{M76} had a big impact on the scientific community by 
demonstrating that simple first-order difference equations exhibit very interesting complex behaviors. 
As a specific example, he illustrated the properties of the logistic 
family of maps $x_{n+1}=\mu x_n(1-x_n)$, where $\mu$ is a real number
parameter between $1$ and $4$, and $x_n$ belongs to the interval $[0,1]$.
This is one  possible generalization of the Ulam--von Neumann map. For 
 $\mu =4$ the recurrence relation
\begin{equation}
x_{n+1}=4 x_n(1-x_n)  \,\,\, \text{$\forall n \geq 0$},
\label{Eq1}
\end{equation}
is said to be (fully) chaotic, meaning that the system is sensitive to the initial conditions (see Ref.~\cite{R98} for a more comprehensive description of the
features of chaos in general and this map in particular). 
In the 1960s, S.M. Ulam and M. Kac~\cite{U60,KU68} provided deeper insight into
this problem by noting that  Eq.~(\ref{Eq1}) has the following  closed--form solution:
\begin{equation}
x_n=\sin^2(\theta \pi 2^n)  \,\,\, \text{$\forall n \geq 0$},
\label{Eq2}
\end{equation}
where now  $\theta=\frac{1}{\pi}\arcsin\sqrt{x_0} \in \mathbb{R}^+$ encodes the initial condition of the system (hence $x_0\mapsto\theta$ is not restricted to the
principal branch of the $\arcsin$ function). 
Let us remark that  
E. Schr\"oder  had already worked out 
the special 
cases $\mu=2$ and $4$ in 1870~\cite{S70}, see also 
Ref.~\cite{TS83}.
We can see that the functional iteration has been reduced to a multiplication and gives the
forward orbit for a given value of $\theta$. We recall that the map involved in Eq.~(\ref{Eq1}) 
is topologically conjugate to the tent map on the interval and
semi--conjugate to the degree two (i.e., doubling) map on the circle~\cite{R98}.
Therefore, if we can solve for one
iterated function, then we can also find the solutions for all topologically conjugate maps. 

\subsubsection{The discrete family of Chebyshev polynomial maps}
The structure of the recurrence relation (\ref{Eq1}) involves polynomials of increasing order: 
$f^{n+1}(x)=4\textrm{T}^2_{2^n}(x)\times f^n(x)$
with $n\in\mathbb{N}$,
$f(x)=4x(1-x)$, $x\in[0,1]$, where $\textrm{T}_2:[-1,1]\rightarrow[-1,1]$, 
$x\mapsto\cos(\beta\arccos x)$,
$\beta\in\mathbb{N}_{>1}$, 
are the Chebyshev polynomials of the first kind~\cite{R74}. Using the transformation $y=(1-x)/2$ we can see that the recurrence relation (\ref{Eq1}) is topologically conjugate to  $\textrm{T}_2$~\cite{GF84}.
With these ideas at hand, we can construct a family of polynomials that map the interval $[0,1]$ onto itself through the recurrence relation: 
\begin{equation}
x_{n+1}=\sin^2\left( \beta \arcsin\sqrt{x_n} \right) \,\,\, \text{$\forall n \geq 0$},
\label{Eq3}
\end{equation}
which has the following closed--form expression (see Ref.~\cite{GF84} for a similar analysis on the interval $[-1,1]$):
\begin{equation}
x_n=\sin^2(\theta \pi \beta^n) \,\,\, \text{$\forall n \geq 0$}.
\label{Eq4}
\end{equation}
For  $\beta=2$,   Eq.~(\ref{Eq4}) corresponds to the logistic recurrence relation (\ref{Eq1}) and its solution is given by Eq.~(\ref{Eq2}). For $\beta=3$, Eq.~(\ref{Eq3}) encodes  the cubic recurrence relation $x_{n+1}= x_{n}(4x_{n}-3)^2$, whose closed--form solution, according to Eq.~(\ref{Eq4}), is $x_n=\sin^2(\theta\pi 3^n)$.
Therefore, Eq.~(\ref{Eq4}) generalizes the analysis carried out by Kac, Ulam and von Neumann~\cite{UvN47,U60,KU68} from $\beta=2$ to any $\beta\in\mathbb{N}_{>1}$. 

\section{Extending the Chebyshev  maps to systems}\label{UvN}

In this paper, we consider any real number $\beta>1$ in Eq.~ (\ref{Eq4}), and we call  ``Chebyshev'' the family of systems obtained from this extension.
Although in general the functions are not polynomials (i.e., for non-integer values of $\beta$)
as it will appear, some defining features of the transformations considered in Ref.~\cite{R74} are present either identically or similarly in the systems we consider,  which justifies the name (we give more justifications below).

\begin{figure}[t]
\begin{center}
\includegraphics[angle=0,width=1.0\linewidth]{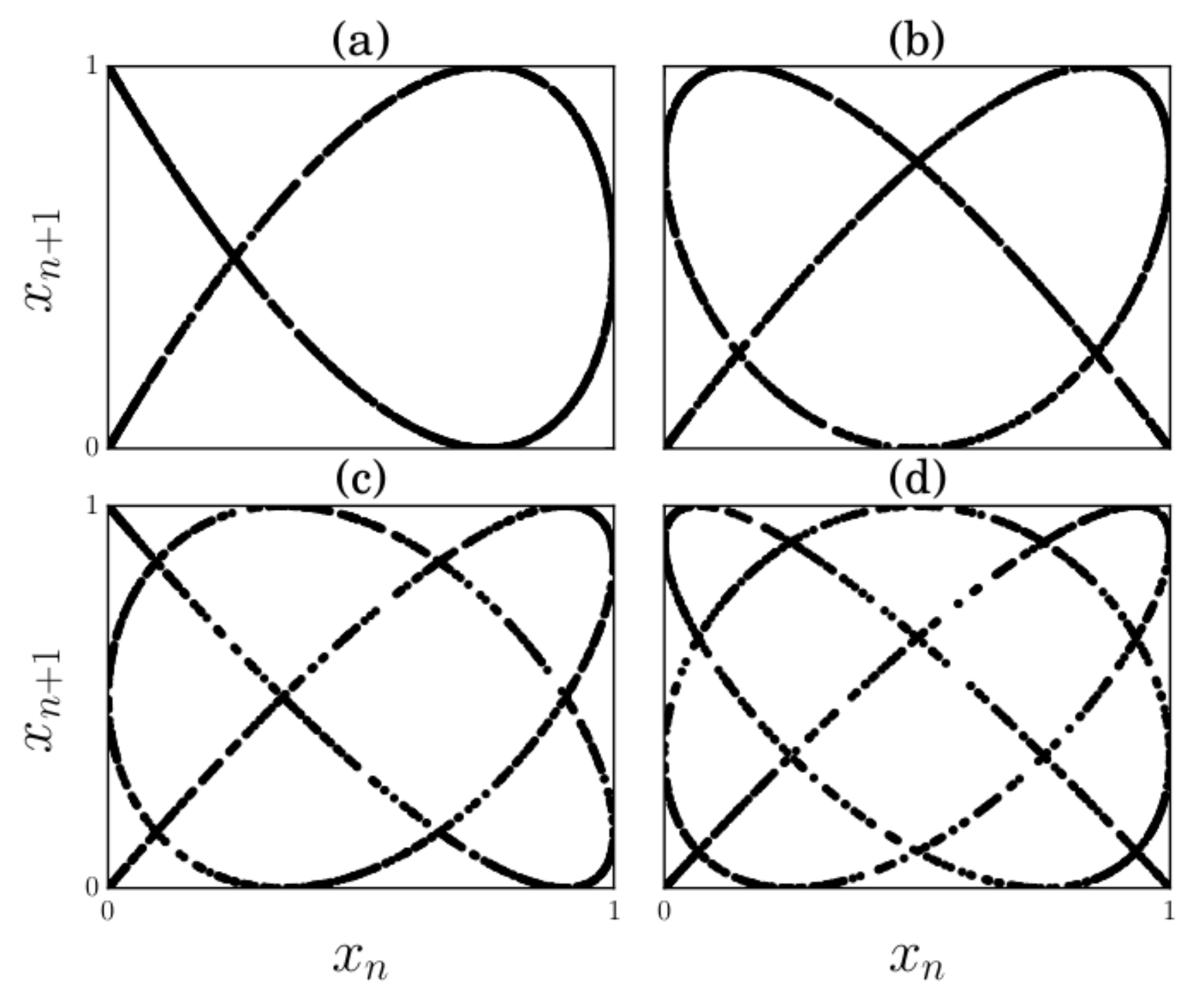}
 \caption{\label{Fig1} Four examples of the graphs produced by Eq.~(\ref{Eq4}) for different $\beta\in\mathbb{Q}_{(1,2)}$  and $\theta=\pi$:
(a) $\beta=3/2;$ (b) $\beta=4/3$; (c) $\beta=5/4$; (d) $\beta=6/5$.}
\end{center}
\end{figure}

In this section we consider only rational values  of $\beta$, and we will consider the irrational cases in Section \ref{irrational}.
First, let us go into the cases when $\beta\in\mathbb{Q}_{(1,2)}$, the set of rational numbers with irreducible representation  $\beta=p/q\in(1,2)$~\cite{G00,T04}.
In order to gain insight into this problem we use graphical displays to uncover the underlying  dynamical features of Eq.~(\ref{Eq4}).
The rules to make these graphs are those of an inverse cobweb plot: 
i)  the values $x_0,x_1,x_2,\ldots$ are produced from Eq.~(\ref{Eq4}), for given values of $\theta$, $p$ and $q$, and 
ii) the ordered pairs $(x_n,x_{n+1})$ are created, forming the points that appear on the plots. 
As shown in Fig.~\ref{Fig1}, the graphs generated exhibit an apparent {\it multivalued} behavior,
i.e., we can observe that  for each $x_n$ it seems that there exists ``simultaneously'' more than one possible value for the next time-step $x_{n+1}$. 
When $\beta=3/2$ (see Fig.~\ref{Fig1}(a)), the graph has two branches and looks like a ``two--valued function''.
For example, if $x_0=3/4$, then $x_{1}$ is equal to  $0$ and $1$.
This characteristic is present for almost all points in the domain $[0,1]$, except when  $x_0=1/4$ or $1$, where the two branches intersect. 
From the other examples shown in Fig.~\ref{Fig1} ($\beta=4/3,5/4$, and $6/5$), we can infer that the number of apparent multivalues of
$x_{n+1}$  depends on  the value of  the denominator $q$. 

The above observation leads us to ask the following question: {\it Can Eq.~(\ref{Eq4}) be conceived as  a class of ``one--to--many'' mappings?} 
In Refs.~\cite{G00} and \cite{T04} this apparent multivalued behavior was related to the emergence of randomness associated with a complete lack of predictability and to the assumption that Eq.~(\ref{Eq4}) cannot be expressed as a map of the type 
\begin{equation}
x_{n+1}=g_n(x_n,x_{n-1},\ldots,x_{n-m+1}), \,\,\, \text{$\forall n \geq m-1$},  
\label{Eq6b}
\end{equation}
for some finite positive integer $m$ (note that in Refs.~\cite{G00} and \cite{T04} it was considered that $g_n=g$ $\forall n$, which is too restrictive to address this issue~\cite{HRS06}).
This interpretation should imply that we could produce some kind of ``truly random sequences'', for example, by programming an algorithm for Eq.~(\ref{Eq4}) on a digital computer.
But, {\it can a digital computer play dices?} 
Obviously not! 
Certainly, the idea of a deterministic mechanism producing such non--deterministic behavior is contradictory. 
In fact the argument presented in Refs.~\cite{G00} and \cite{T04} of a {\it lack of exact predictability} (i.e., the nonexistence of maps $g_n$ as in Eq.~(\ref{Eq6b}) to produce the graphs shown, for example, in Fig.~\ref{Fig1}), is wrong, leading to some flawed conclusions as was stated in Ref.~\cite{M04}, because such maps do exist (see Eq.~(\ref{Eq6aa}) below).

\subsection{Function systems description}\label{SecIFS}

\begin{figure}[t]
\begin{center}
\includegraphics[angle=0,width=1.0\linewidth]{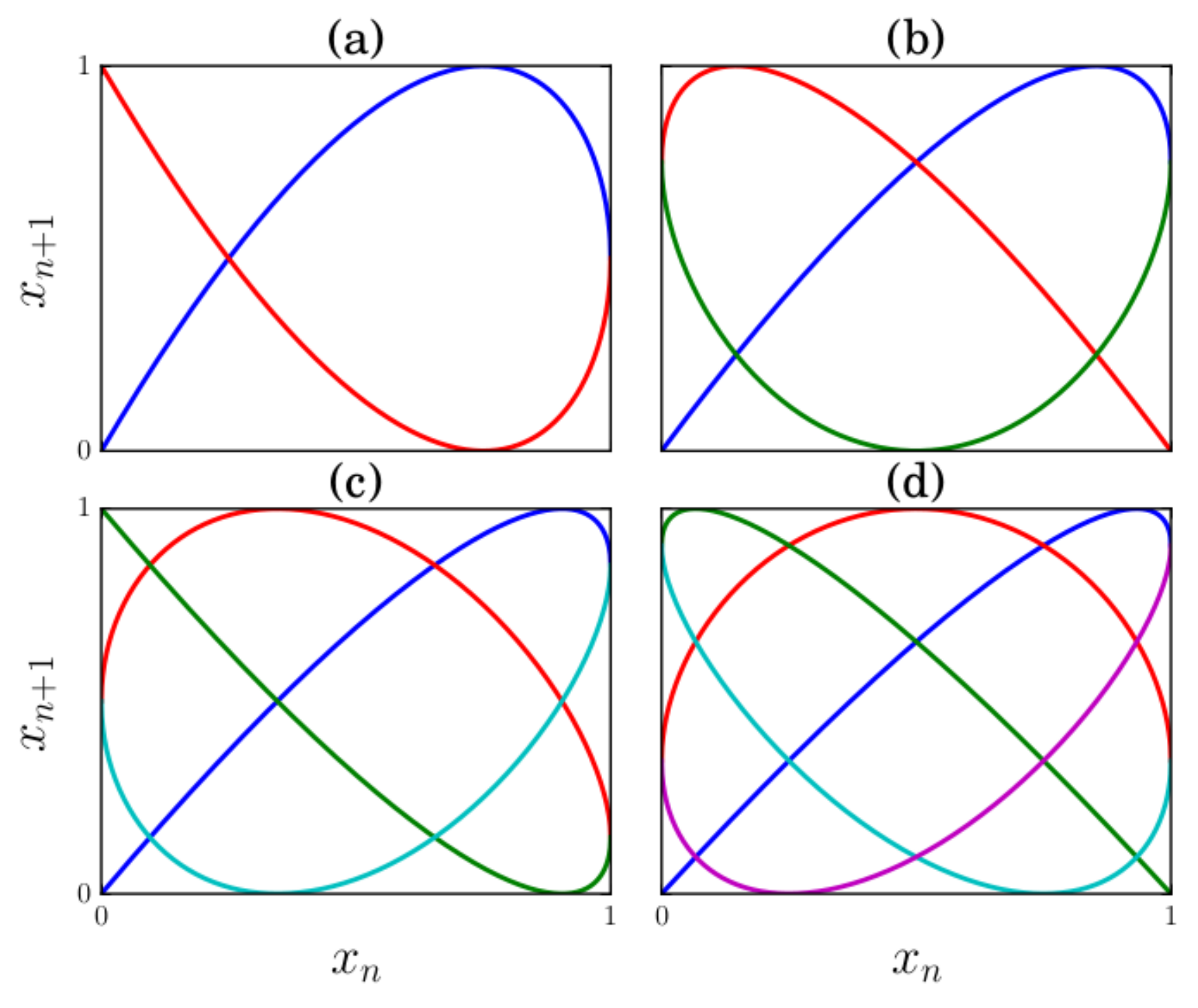}
 \caption{\label{Fig2} 
 Four examples of function systems  calculated with Eq.~(\ref{Eq6aa}) for different $\beta=p/q$ as in Fig.~\ref{Fig1}.
The colors are associated with the value of $\omega$ in Eq.~(\ref{Eq6aa}): $\omega=0$ (blue); $\omega=1$ (red); $\omega=2$ (green);
$\omega=3$ (cyan); $\omega=4$ (violet).
}
\end{center}
\end{figure}

Now we proceed to elucidate a possible mathematical structure for the correct understanding of these apparent multivalued dynamics.
To fix the idea, consider the case $\beta=3/2$ and note that the function $f_0(x):=\sin^2\left( \frac{3}{2}\arcsin\sqrt{x} \right)$
reproduces the ``upper branch'' of Fig.~\ref{Fig1}(a), which starts at point $(0,0)$, has a maximum at $(3/4,1)$, and ends up at $(1,1/2)$. 
On the other hand, the function $f_1(x):=\sin^2\left( \frac{3}{2}(\arcsin\sqrt{x} +\pi )\right)$
reproduces the ``lower branch'' of Fig.~\ref{Fig1}(a), which starts at point $(0,1)$, has a minimum at $(3/4,0)$, and ends up at $(1,1/2)$. Both branches intersect at $(1/4,1/2)$ and meet together at $(1,1/2)$. 
In Fig.~\ref{Fig2}(a),  the blue and red lines correspond to $f_0$ and $f_1$, respectively. 
The sequence generated from $x_n=\sin^2(\theta\pi (3/2)^n)$ resembles  a ``random selection''~\cite{A98} between the functions $f_0$ and $f_1$, i.e., these functions define the set $\{f_0(x),f_1(x)\}$ of possible outcomes for each $x\in[0,1]$.

More generally, we can verify that for any  $\beta=p/q$, Eq.~(\ref{Eq4}) is the closed--form solution of the system $x_{n+1}  = f_{\omega_n}(x_n)$ for all $n\in \mathbb{N}$ given by
\begin{equation}\label{Eq6aa}
x_{n+1}  = \sin^2\left(\beta (\arcsin\sqrt{x_n}+ \omega_n\pi) \right),
\end{equation}
 with $x_0\in[0,1]$ and where
$\{ \omega_n \}_{n \in \mathbb{N}}$ a piecewise increasing sequence of integers  taking values in $\mathbb{Z}_q := \{0,1,\ldots,q-1\}$ such that
\[
 \omega_n = \lfloor \theta \beta^n \rfloor \bmod q \, \text{ if } \{ \theta \beta ^n \} \in [0,1/2) \, \text{ and } \, \omega_n = \lceil \theta \beta ^n \rceil \bmod q \, \text{ otherwise,}
\]
where $\{ \cdot \}$ and $\lfloor \cdot \rfloor$ denote the fractional and integer part, respectively.
Now $\theta$ is any positive real number such that
\begin{eqnarray}\label{Branches}
 \frac{1}{\pi} \arcsin \sqrt{x_0} \; = \left\{
 \begin{array}{cl}
 \{ \theta \} 		& \text{if } \,	\{\theta\} \in [0,1/2) \\
 1 - \{ \theta \}  	& \text{if } \,	\{ \theta \}  \in [1/2,1) 
 \end{array} \right.    \in [0, 1/2 ],
\end{eqnarray}
where the principal branch of the inverse sine function applies.
Note that to a given $x_0 \in [0,1]$ corresponds infinitely (countably) many values of $\theta$ in $\mathbb{R}^+$, and that  to two different $\theta \in \mathbb{R}^+$ correspond different sequences $\{\omega_n\}_{n \in \mathbb{N}}$.
Therefore, unless we restrict $\theta\in[0,1/2]$ which ensures a one-to-one correspondence with $x_0$, the knowledge of $x_0$ is not enough to anticipate the sequence of functions that will be used along the dynamics.
To this purpose, we also need to know the first function employed, for which we need to know $\omega_0$. 
(A consequence of this will be explored below, c.f. Fig.~\ref{Fig3}).
In Fig.~\ref{Fig2}, we show the characteristic branches  of the
graphs created from Eq.~(\ref{Eq6aa}) for four different values of the parameter $\beta=p/q$, such that $1<\beta<2$, confirming  indeed that the
value of $q$ determines the number of functions that are needed to construct  the set $\mathcal{F}_\beta := \{ f_0,f_1,\ldots,f_{q-1} \}$.
Observe that independently of $1<\beta<2$ each function is unimodal. More generally, for any $\beta>1$ each function in the system $\mathcal{F}_\beta$
is $(\lceil \beta \rceil -1)$-modal.
Hence the choice of naming ``Chebyshev'' these systems. We can prove that the {\it function system} $\mathcal{F}_\beta$ has finitely many functions if and  only if $\beta$ is a rational
number, and for each such $\beta=p/q$, with $1\leq q<p$ coprime integers, $\mathcal{F}_{p/q}$ has exactly $q$ functions~\cite{MT17}.

What is interesting so far is that we found a one--parameter family of chaotic systems, where the maps (\ref{Eq1}) and (\ref{Eq3}), that is  when $\beta\in\mathbb{N}_{>1}$, are particular cases of Eq.~(\ref{Eq6aa}).
In textbooks on dynamical systems, chaos is commonly exemplified by the logistic map (\ref{Eq1}), and the iteration of Chebyshev maps ($\beta\in\mathbb{N}_{>1}$) which generate mixing transformations that model canonical features of chaos~\cite{AR64}.
Therefore, a promising research program in nonlinear science can arise around the following question:
{\it What is the physical interpretation and the consequences of this mathematical construct?}

\subsection{Statistics on switchings among maps in $\mathcal{F}_\beta$}\label{Switching}
\begin{figure}[!ht]
\begin{center}
\includegraphics[angle=0,width=1.0\linewidth]{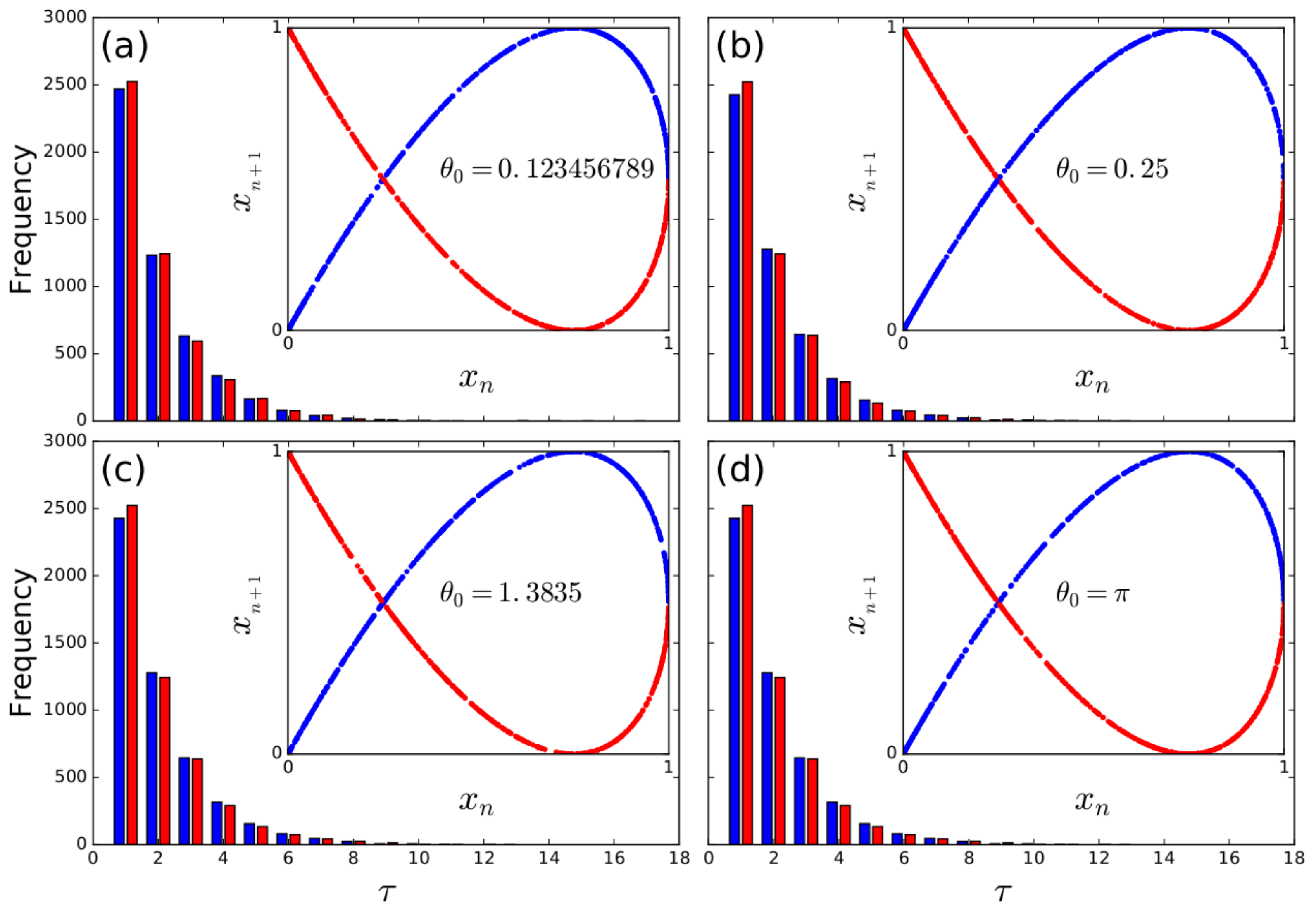}
 \caption{\label{Fig3a} 
 Number $\tau$ of consecutive usage of the same branch, upper ($0$) or lower ($1$), calculated in time steps
with Eq.~(\ref{Eq4}~$\sim$ \ref{Eq6aa}) 
for $\beta=3/2$ and four different 
initial conditions $\theta_0$.
The total number of points is  $N=20\times 10^3$.
The colors are associated with the upper branch (red) and the lower branch (blue). Only statistics on words $0^{\tau}$ and $1^{\tau}$, $\tau\geq 1$, are
shown. 
}
\end{center}
\end{figure}

\begin{figure}[!ht]
\begin{center}
\includegraphics[angle=0,width=1.0\linewidth]{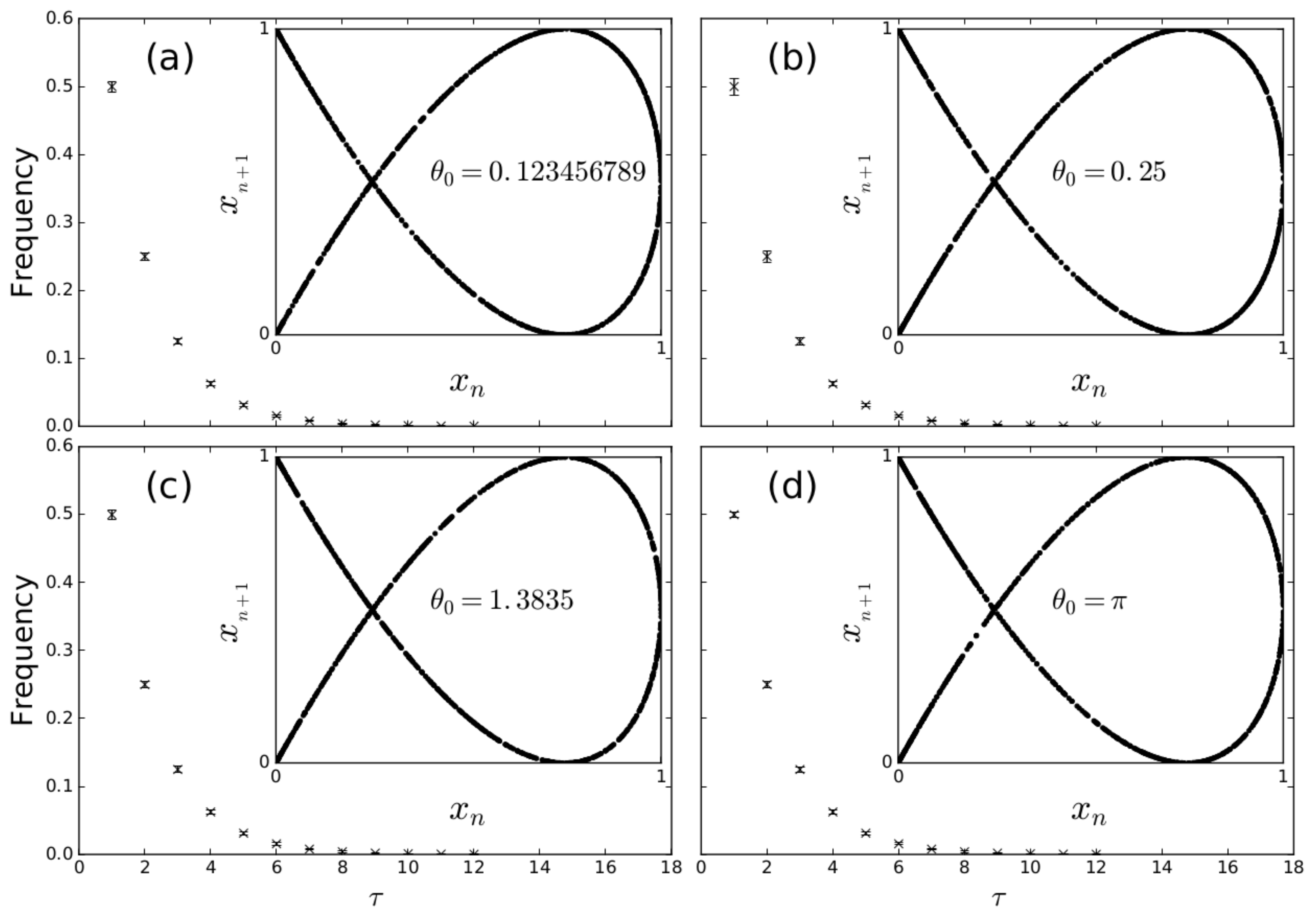}
\caption{
More statistics for $\beta = 3/2$.
 Frequency mean and deviation to the mean among all words in $\{0,1\}^\tau$ for each given length $1 \leq \tau \leq 12$. 
 For instance, at $\tau = 2$, all words in $\{00, 01, 10, 11\}$ are considered. 
 (Note that only the words $00$ and $11$ are shown in Fig.~\ref{Fig3a}.)
 Four different initial conditions are shown, each for a total number of points $N=20 \times 10^3$ using Eq.~(\ref{Eq4}~$\sim$ \ref{Eq6aa}).
}
\label{WF32}
\end{center}
\end{figure}

\begin{figure}[!ht]
\begin{center}
\includegraphics[angle=0,width=1.0\linewidth]{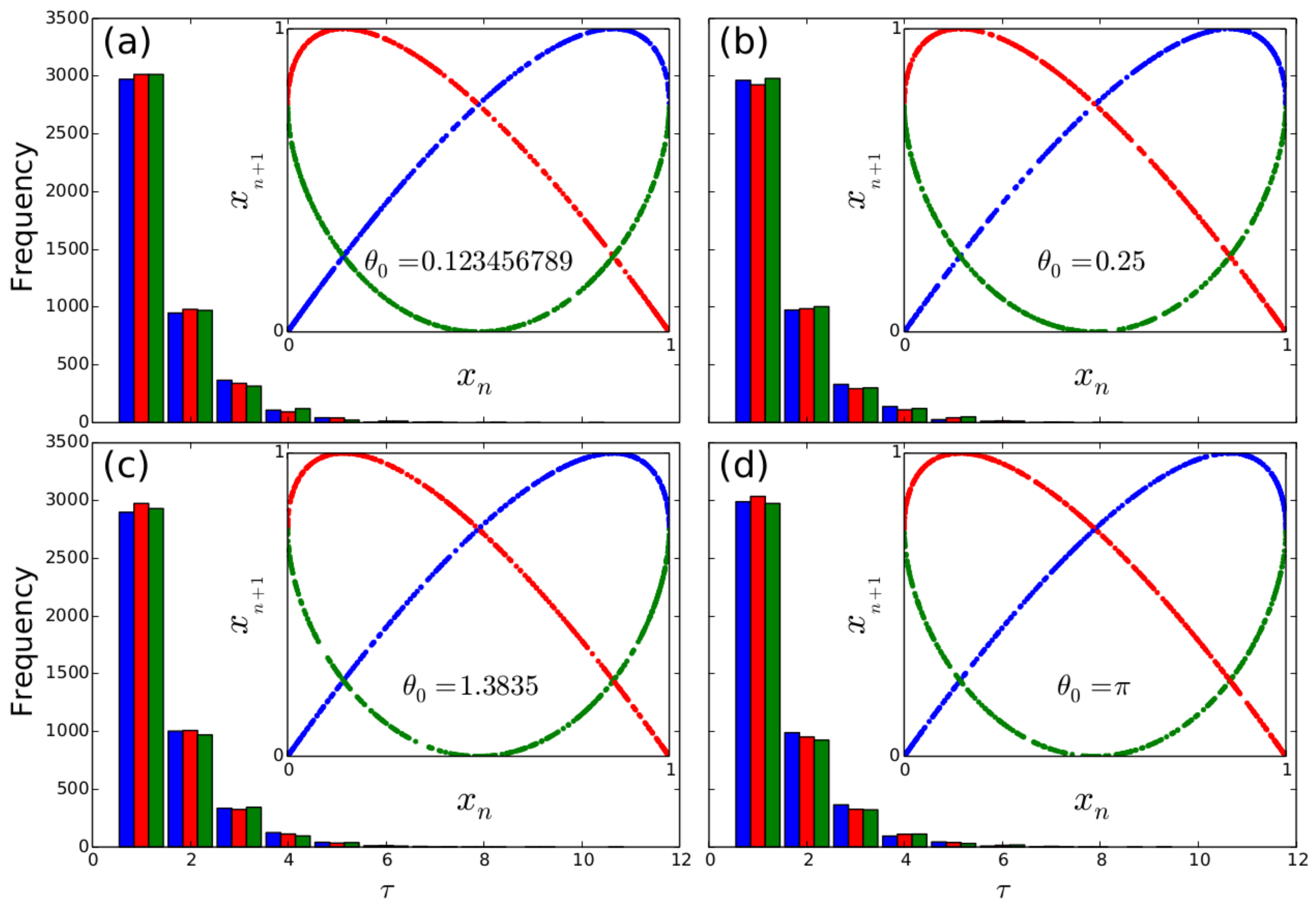}
\caption{
Statistics for $\beta = 4/3$.
Number $\tau$ (permanence time) of consecutive usage of the same branch, $\omega=0$ (red), $\omega=1$ (green), or $\omega=2$ (blue), calculated in time steps with Eq.~(4), for four different initial conditions $\theta_0$. 
The total number of points is $N = 20 \times 10^3$ using Eq.~(4). 
Only statistics on words $\omega^\tau$ with $\tau \geq 1$ are shown.
}  
\label{TR43}
\end{center}
\end{figure}

\begin{figure}[!ht]
\begin{center}
\includegraphics[angle=0,width=1.0\linewidth]{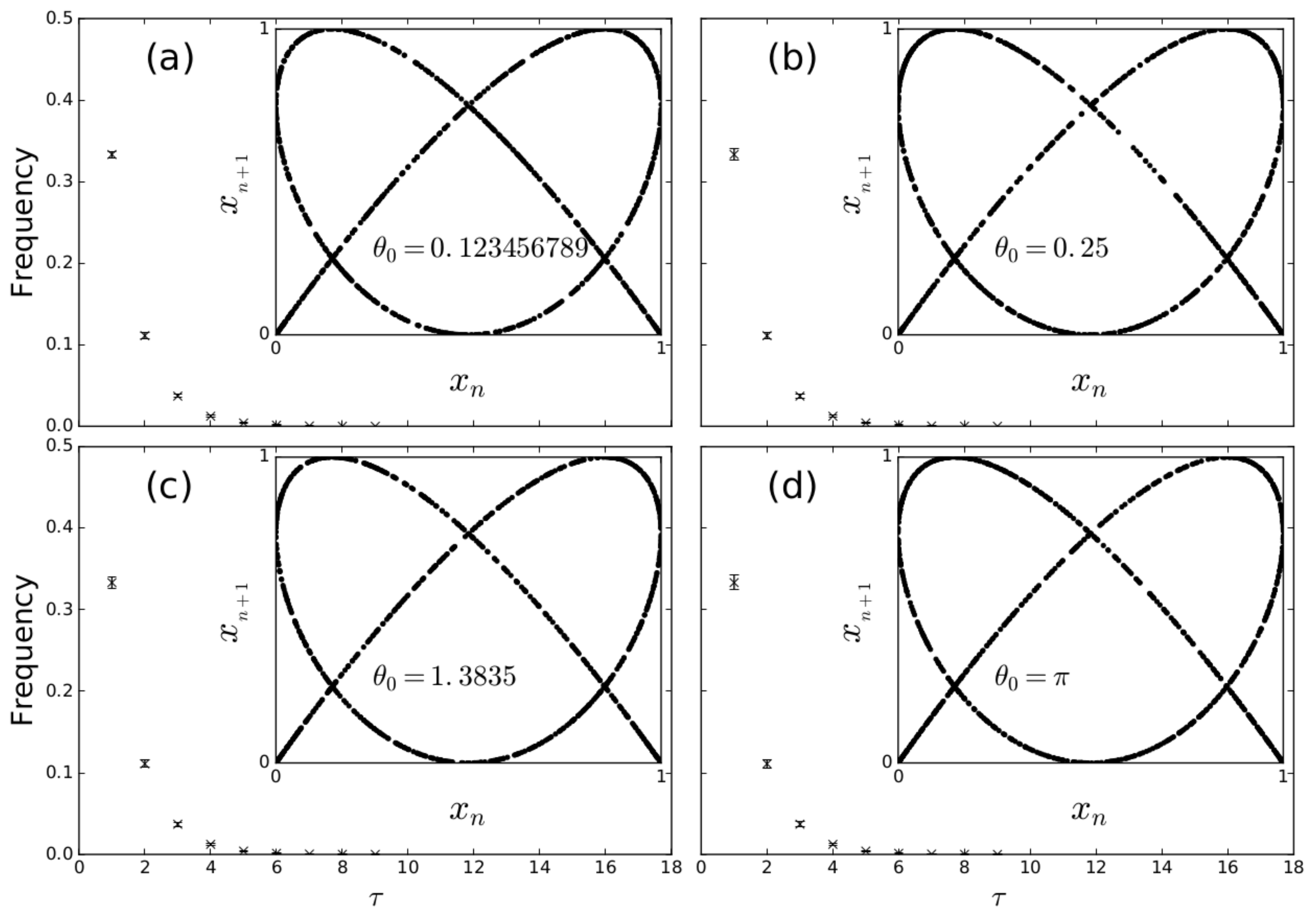}
\caption{
More statistics for $\beta = 4/3$.
 Frequency mean and deviation to the mean among all words in $\{0,1,2\}^\tau$ for each given length $1 \leq \tau \leq 9$. 
 Therefore there are $3^\tau$ possible words for each $\tau \geq 1$. 
 (Note that only the words $000$, $111$ and $222$ are shown in Fig.~\ref{TR43}.)
 Four different initial conditions are shown, each for a total number of points $N=20 \times 10^3$ using Eq.~(4).
}
\label{WF43}
\end{center}
\end{figure}

We have found a mathematical setting to interpret Eq.~(\ref{Eq4}) as the closed--form expression of the orbits of the discrete dynamics Eq.~(\ref{Eq6aa}) using a function system,
which is a  collection of maps such that at each time step a map $f_{\omega}$ is chosen in an independent fashion.
In Figs. \ref{Fig3a} and \ref{WF32}, we show  numerical evidence that for $\beta =3/2$ the choice is identically distributed with statistics over a single typical orbit.
By ``typical'' we mean with respect to an invariant  measure which is absolutely continuous with respect to Lebesgue and has density $\rho(x)=\frac{1}{\pi \sqrt{x(1-x)}}$ 
which does not depend on $\beta>1$~\footnote{A numerical evidence has also been reported in~\cite{condmat06}, where statistics were performed over single orbits.
A proof when $\mathcal{F}_{\beta}$ is a singleton (i.e. when $\beta$ is an integer) can be essentially found in \cite{AR64}, but does not work when $\beta$ is not an integer.
However, this results when
$\beta$ is not an integer  can be established showing that $\rho$ is a fixed point of some transfer operator. This will be published elsewhere~\cite{MT17}.}.

\begin{figure}[!ht]
\begin{center}
\includegraphics[angle=0,width=1.0\linewidth]{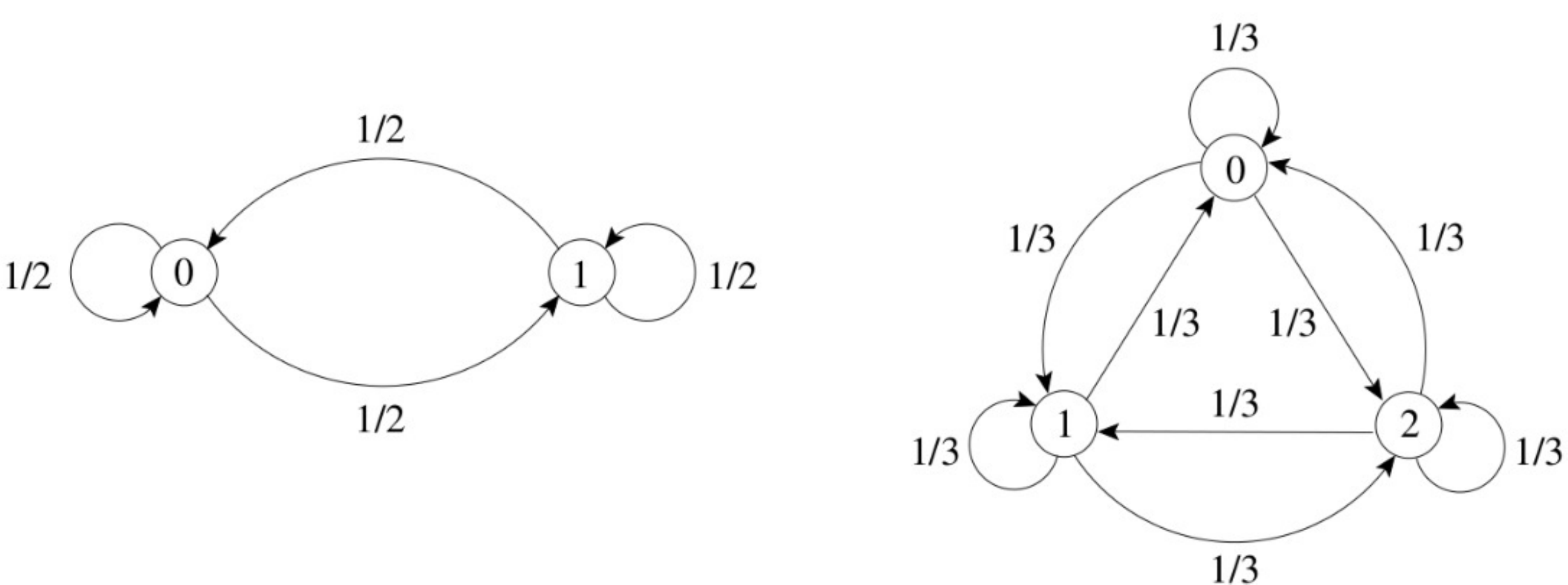}
\caption{
Examples of the Markov chain model for two choices among maps in $\mathcal{F}_\beta$ for $\beta=3/2$  (left, corresponding to Figs.~\ref{Fig3a} and \ref{WF32})  and $\beta = 4/3$ (right, corresponding to Figs.~\ref{TR43} and \ref{WF43}). 
Observe that for any $\beta = p/q$, $1\leq q < p$ coprime, the probability transitions are all given by $1/q$.}
\label{FullShifts}
\end{center}
\end{figure}

In Figs. \ref{TR43} and \ref{WF43} we show similar statistics for the case when $\beta = 4/3$, 
indicating   that the choices are also identically distributed. More cases were studied,  arriving also to 
the same conclusion (data not shown).
This numerical analysis suggests that
for  any $\beta=p/q$, with $1\leq q <p$ coprime, we have that $\mathbb{P}(u \in \{0,1,\ldots,q-1\}^\tau | u \in \omega(\theta_0)) \sim 1/q^\tau$.
A model to reproduce this statistics is given by a full-shift on $q$ symbols. 
In Fig.~\ref{FullShifts} we show for $\beta = 3/2$ and $4/3$ the corresponding  Markov chains, 
meaning that any path obtained from these graphs corresponds to an initial condition in $[0,1]$ that produces this sequence of choices among the maps in $\mathcal{F}_\beta$, with the same statistics as reported in  Figs.~\ref{Fig3a} to \ref{WF43}.

\subsection{Bi--sensitivity to initial conditions}\label{BiSen}
\begin{figure}[t]
\begin{center}
 \includegraphics[angle=0,width=1.05\linewidth]{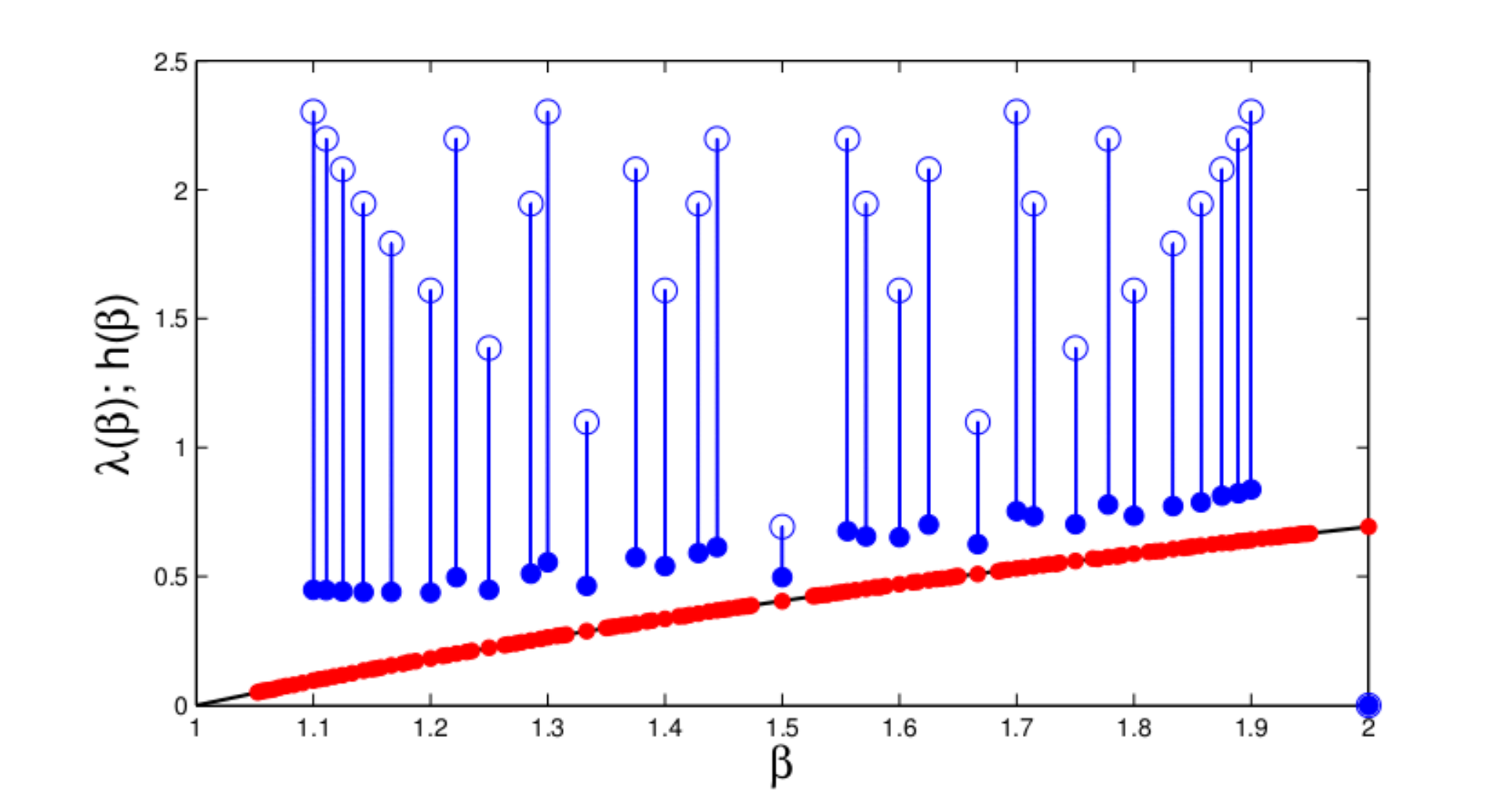}
 \caption{\label{Fig4} 
Numerical estimation of complexity.
Red dots correspond to the Lyapunov exponent $\lambda(\beta)$  
with $126$ rational values of $\beta=p/q \in (1,2]$, distributed according to the Farey sequence  (with depth $20$) using Eq.~(\ref{Eq6aa}) with $N=1\times 10^3$ (the black curve is $\log\beta$). 
The vertical blue segments represent the range of possible values calculated for the combinatorial entropy $h(\beta)$ of switches between functions, with theoretical upper bound $\overline{h}(p/q)=\log q$ (blue circles) and numerical estimate of $h(\beta)$ (blue dots).
For the latter calculation $N=200\times 10^3$ and a words length of $11$ was used for $31$ rational values of $\beta$ (up to depth $10$ in the Farey's sequence).
} 
\end{center}
\end{figure}

An essential feature of chaotic dynamical systems is the {\it sensitive dependence on initial conditions}. 
If the initial condition is only approximately specified, then the evolution of  two nearby approximating  such  states may be very different. 
What we have discovered so far is that the Chebyshev dynamical systems are not only sensitive to initial conditions, but also, when $\beta\notin\mathbb{N}_{>1}$ we have at every time step a sensitive choice among several evolution maps. 
It turns out that the sequence of selections between the different functions is partially encoded in the initial condition $x_0$ (see Eq.~(\ref{Eq6aa})), while it appears to be totally encoded when $\beta\in\mathbb{N}_{>1}$ (for which the function system reduces to a singleton, a Chebyshev polynomial map). 
While in the usual sense of chaos, the evolution law is unique and the ``uncertainty'' depends only on the limited information we have on the initial conditions, in our case the evolution rule is moreover selected among a collection  of maps $\mathcal{F}_\beta$ typically with a nontrivial combinatorics that also depends on the initial condition $\theta$, of which $x_0$ is only a partial observation, c.f. Eq.~(\ref{Branches}). 
Hence, the concept of a {\it self--ruling system}.
We propose to qualify such kind of chaotic dynamics as a class of dynamical systems with {\it bi--sensitivity} to initial conditions.

The manifestation in the dynamics of this bi--sensitivity, although observable in the apparent ``multivaluedness'' (see Fig.~\ref{Fig1}) may not, however, be easily quantified. 
Indeed, Fig.~\ref{Fig4} shows (red points) a numerical estimation of the Lyapunov exponent 
$$\lambda_\beta(x_0):=\limsup_{n\to\infty}\frac{1}{n}\sum_{j=0}^{n-1}\log(|f'_{\omega_j}(x_j)|),$$
for several values of $\beta \in \mathbb{Q}_{(1,2]}$ using Eq.~(\ref{Eq6aa}), and a typical value of $x_0$.  
It suggests that the Lyapunov exponent $\lambda_\beta$ is a continuous function of $\beta$ over the rationals, and indeed we can prove~\cite{MT17}
that for every $\beta>1$,  $\lambda_\beta=\log\beta$ for almost all $x_0 \in [0,1]$. 
This interpolates the obtained Lyapunov exponents for those cases when $\beta \in \mathbb{N}_{>1}$ \cite{G00}. 
However, this quantity is not a faithful estimate of the complexity of the system for  $\beta\notin\mathbb{N}_{>1}$, since it does not give information on the sequence of maps used along the orbit. 
The combinatorial entropy $h(\beta)$ \cite[p. 340]{R98} of this sequence $\{\omega_n\}_{n \in \mathbb{N}}$ for the same $\beta$ values can also be estimated as shown in Fig.~\ref{Fig4} (blue segments).
The upper circles correspond to the theoretical maximal possible combinatorial entropy $\overline{h}_\beta$ corresponding to the full-shift on $q$ symbols introduced in Section \ref{Switching}. 
The function $\overline{h}_\beta = \log q$ is not a continuous function of $\beta$ anywhere over
$\mathbb{Q}_{(1,2)}$ (observe that $\overline h_\beta =0$ for every $\beta \in \mathbb{N}_{>1}$). Note that using a different approach, $\lambda_\beta + \overline{h}_\beta$ for every $\beta>1$ was obtained in Ref.~\cite{G00} as an estimate of the topological entropy of the system. 
The lower dots are numerical estimations of  ${h}(\beta)$. 
Hence, for a given $\beta \in (1,2]$ the segment $[h(\beta),\overline{h}_\beta]$ between the lower dot and the upper circle gives 
the numerical limitation on our estimation of combinatorial entropy.
However, all of our statistics presented in Figs.~\ref{Fig3a} to \ref{WF43} indicate that the corresponding blue dot in Fig. \ref{Fig4}  should be within the corresponding blue circle (i.e., any vertical segment reduces to a point).
This fact is trivially valid for $\beta=2$.

Although a more optimal numerical estimation may be found, this result already clarifies the source of the complexity in these systems and indicates the reason of the flawed conclusions drawn in Refs.~\cite{G00} and \cite{T04}.
For instance, although $\beta <2$, higher ``entropies'' than $\log 2$ can be found in this family.
Also, numerical estimations indicate that $h(\beta) \to \infty$ as $q \to \infty$, hence Diophantine approximations of irrational numbers tend to an infinite (upper bound) entropy. We will come back on these irrational values for $\beta$ in Section \ref{irrational}.

 \begin{figure}[t]
\begin{center}
\includegraphics[angle=0,width=1.0\linewidth]{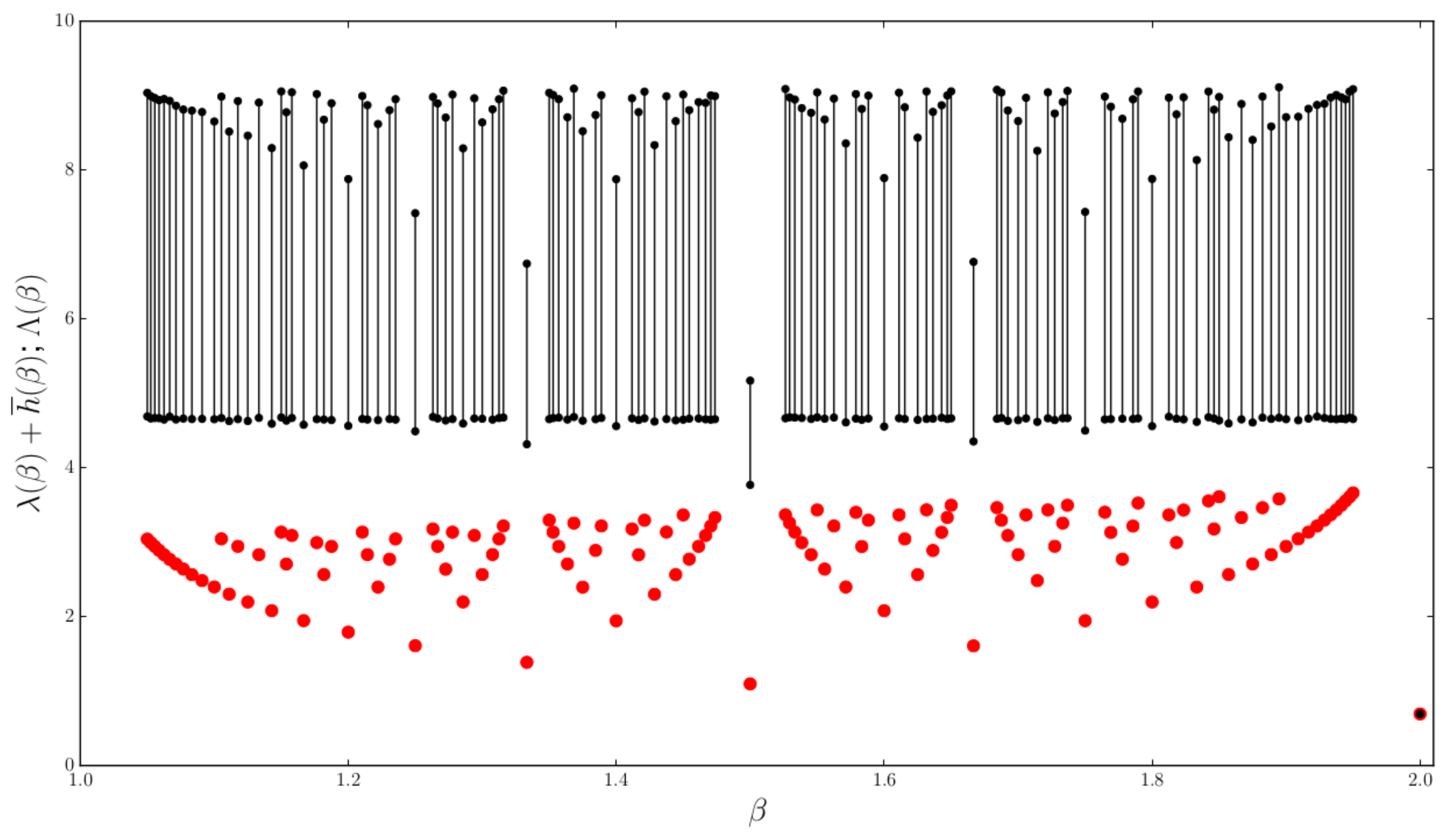}
\caption{
Comparison of Lyapunov exponent estimates between TISEAN (upper black dots) and our results from Fig.~\ref{Fig4}  (lower red dots).
The parameters used for TISEAN are $\epsilon = 10^{-3}$ (default), embedding dimension  = 1, and the first 20 iterates of each series have been compared.
}  
\label{COMPLEXITY}
\end{center}
\end{figure}

Also we show that when applying proved methods of Lyapunov exponent estimation for continuous dynamical systems (i.e.,  function systems as ours but with only one continuous function), results can strongly differ from our estimates reported in Fig.~\ref{Fig4}.
We present a similar plot in Fig.~\ref{COMPLEXITY} with an estimation of Lyapunov exponents $\Lambda(\beta)$ using the TISEAN package \cite{H99}, shown as the upper series of black segments.
For each segment, the upper dot corresponds to a linear interpolation with the two left-most points before the plateaus, while the lower dot corresponds to a linear interpolation with the three left-most points.
The lower series of red dots reproduces the $\lambda_\beta + \overline{h}_\beta$ complexity function of $\beta \in (1,2]$, shown in Fig.~\ref{Fig4}.
We can see that the TISEAN package systematically gives higher values than our upper estimation of the complexity, excepted at $\beta =2$ where they coincide as expected (where the red dot is seen within the black dot), and where the function system is indeed a singleton.
For $\beta \in (1,2)$, the global shape of the distributions of the upper black dots and lower red dots are, however, strikingly similar between the two estimates.

The discrepancy between the two estimates can be explained by the possibility, when there are more than one function in the function system, of an ``instantaneous'' separation of arbitrarily close states without more stringent bounds than the diameter of the state space.
The red dots in the lower part of Fig.~\ref{Fig4} are however more reliable in estimating the complexity, since it considers explicitly the switching mechanism and  arbitrarily close initial conditions.

\subsection{Branching and coalescence}\label{Coala}
The future dynamics  generated from Eq.~(\ref{Eq6aa}) depends on the past in a nontrivial manner.
If we look at the values generated by Eq.~(\ref{Eq4}) as an ``experimental time--series'';
{\it then how difficult  is it to perform computations with full or partial knowledge of the past values of the series
to predict its outcomes over time?} This  seems difficult because there exists an intrinsic limitation
on the possibility to reconstruct the process if we do not know the choices between the  functions in the collection 
$\{ f_0,f_1,\ldots,f_{q-1} \}$
at each time step.
This leads us to 
 the idea of {\it branching}~\cite{S07}:  Among all admissible real valued sequences $\{x_i\}_{i\in \mathbb{N}}$,
 there exist two sequences $x$ and $x'$ and instants $0\leq n \leq m$ such that
$x_n,x_{n+1},\ldots,x_{m}=x_{n}',x_{n+1}',\ldots,x_{m}'$ and $x_{m+1}\neq x'_{m+1}$ or $x_{n-1}\neq x'_{n-1}$. 
We also say that there is {\it coalescence} over the time window $[n,m]$.
We recall that this situation cannot happen in  usual dynamical systems defined with a single map~\cite{R98}.

\begin{figure}
\begin{center}
 \includegraphics[height=0.2cm,angle=0,width=1.0\linewidth]{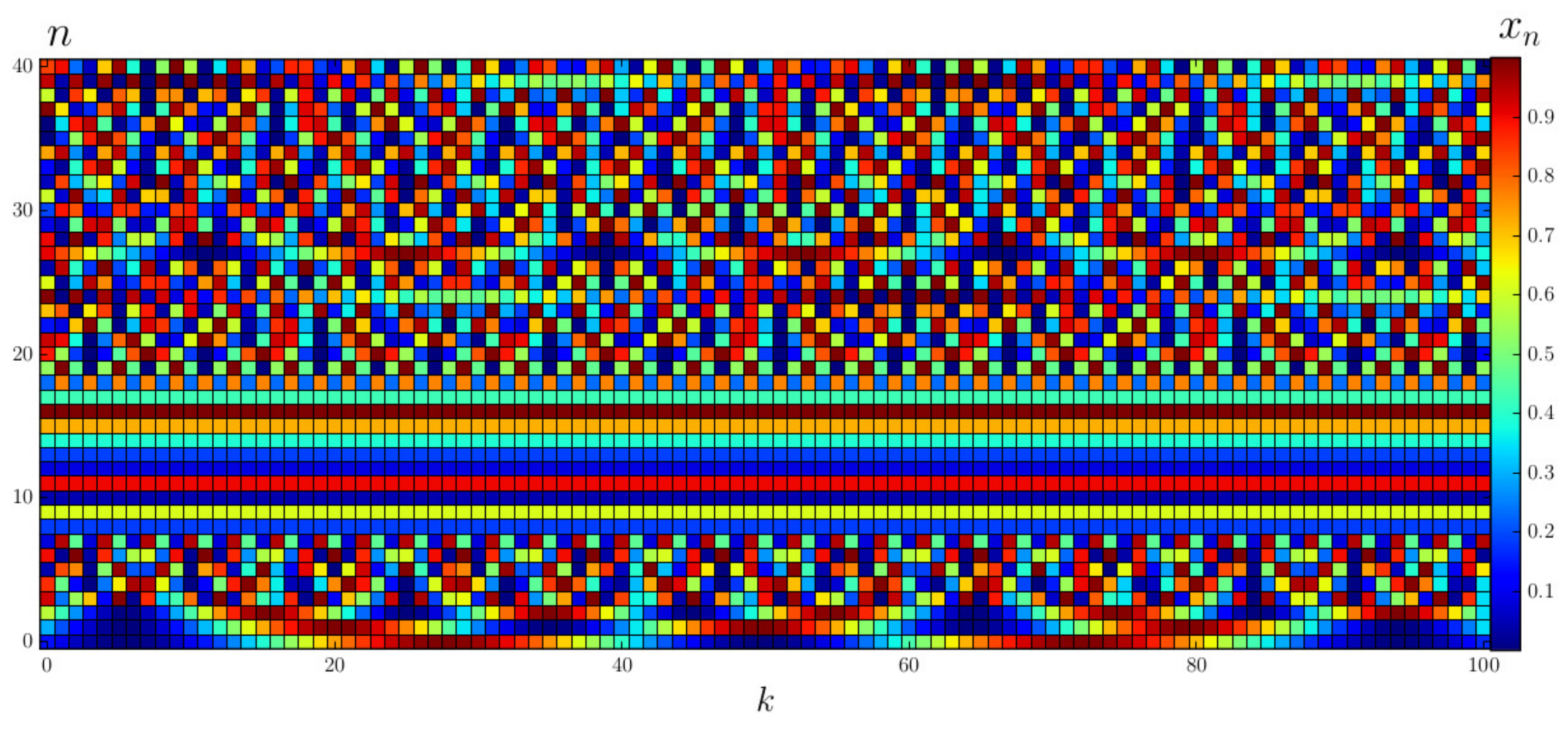} \includegraphics[angle=0,width=1.0\linewidth]{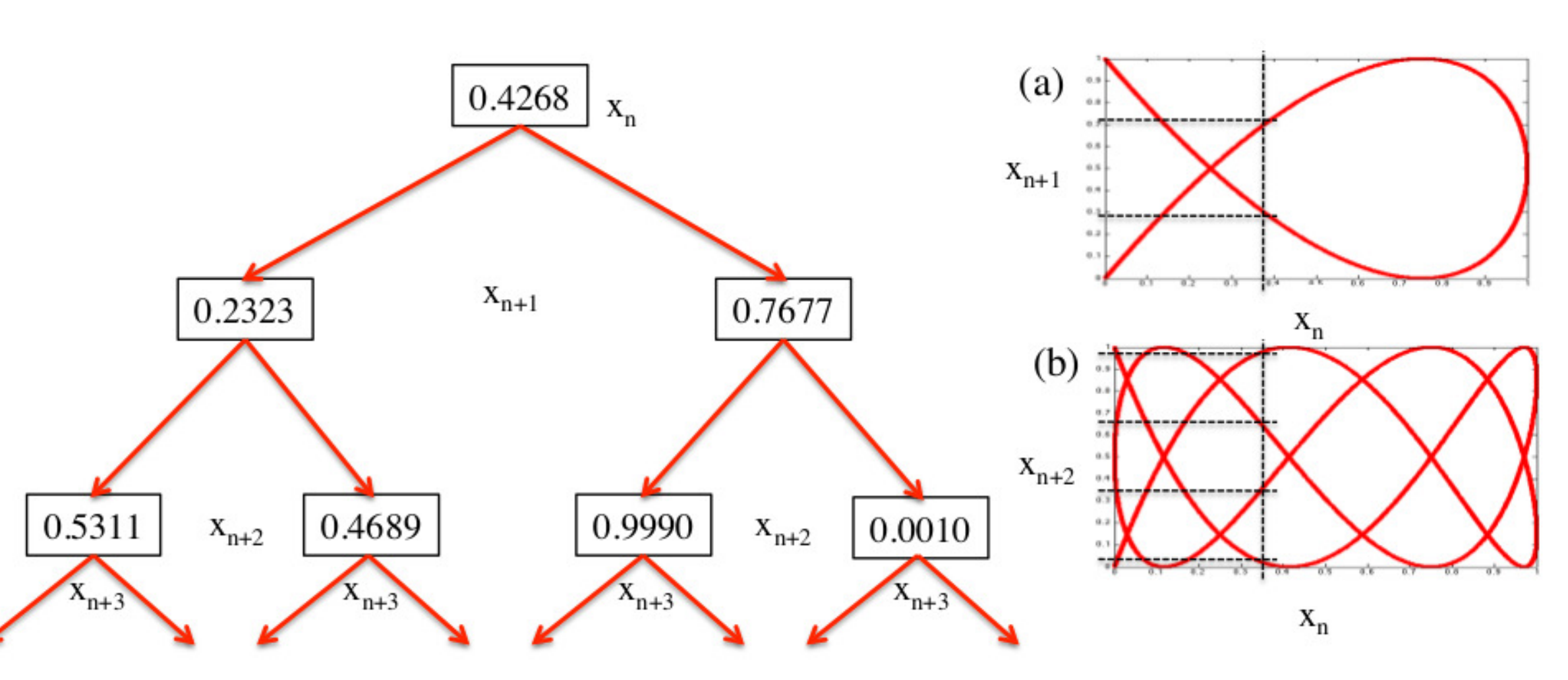}

 \caption{\label{Fig3} 
Upper figure: Coalescence and branching of values  
$x_n^{(k,m,s)}$ after Eq.~(\ref{Eq7})  
for $\beta=3/2$, with $k\in[0,100]$, $m=9$, $s=8$, and $\theta_0=\pi$. The color of each tile in the mosaic
represents the value of $x^{(k,9,8)}_n\in[0,1]$. Lower figure: Example of branching forward in time starting from $x^{(k,9,8)}_{17}$ and the graphs 
produced with Eq.~(\ref{Eq6aa}) for $\beta=3/2$: (a) $x_{n+1}$ vs $x_n$ and (b) $x_{n+2}$ vs $x_n$.} 
\end{center}
\end{figure}

In the following, we explain how {\it uncertainty} can emerge from a branching process and
 be emulated from the dynamics generated by Eq.~(\ref{Eq6aa}).
To do so we define the following family of sequences:
\begin{equation}
x_n^{(k,m,s)}:=\sin^2\left( \pi\left( \theta_0+q^mk \right) \left(q/p\right)^s\left(p/q\right)^n \right), 
\label{Eq7}
\end{equation}
 $\forall n \geq 0$, where $k,m,s\in \mathbb{N}$ and $\theta_0\in\mathbb{R}^+$. 
 Note that we have discretely  distributed $\theta$ in Eq.~(\ref{Eq4}) using the set $\left\{(\theta_0+q^mk ) \left(q/p\right)^s\right\}_{k,m,s\in \mathbb{N}}$. 
Here  $k$ parameterizes the different sequences, while the values 
$\{x^{(k,m,s)}_n\}_{n=s}^{s+m}$ are the same for any $k$, i.e., $x_n^{(k,m,s)}=\sin^2 \left( \pi\theta_0\left( p/q \right)^{n-s}\right)$
 for all $s\leq n\leq m+s$,
 because $\{kp^{n-s}q^{m-n+s}\}_{k,m,s\in\mathbb{N}} \in\mathbb{N}$.
However, a branching occurs at $n=s+m$, because $x_{s+m+1}^{(k,m,s)}=\sin^2\left(  \pi\theta_0\left( p/q \right)^{m+1} +\pi k p^{m+1}/q \right)$,
for $x_{s+m+1}^{(k,m,s)}$ has $q$ different possible outcomes. 
Although within the coalescence window we cannot distinguish which $k$ corresponds to the value of $x_{n}^{(k,s,m)}$ (since all those states degenerate to a single value), at $n = s + m + 1$ we can distinguish from the value of the state (only $q$ of them) whether $k$ is $q-1, q-2 \ldots$ or $0$ in $\mathbb{Z}_q$.
The reasoning can be pursued by induction so that from a long enough observation of the future beyond the coalescence window one could determine which $k$ stands for the orbit $\{ x^{(k,m,s)}_{n} \}_{n\geq s+m}$, hence the whole orbit too starting from $n=0$.
Therefore, we say that the system is \emph{transitorily uncertain}.
Likewise,  the state $x_{s-1}^{(k,m,s)}=\sin^2\left(  \pi\theta_0\left( \frac{q}{p} \right) +\pi k q^{m+1}/p \right)$
has $p$ possible values. Now, for given $\beta>1$ and $\theta_0 \in \mathbb{R}^+$, $s$ and $m$ can be as large as one wishes and therefore the number of initial conditions for which transitory uncertainty eventually happens is countable.
As there are uncountably many values for $\theta$, we conclude that it is so for initial conditions for which transitory uncertainty eventually happens.
Finally, independently of the size of the coalescence window, from which the transitory uncertainty originates,
 we can approximate true uncertainty by taking the limit $q \to \infty$ and keeping $\beta\in(1,2)$, so that our argument is complete.
 The latter corresponds to take irrational values for $\beta$, which we further consider in Section 4.
However, note that the combined effect of this transitory uncertainty with the ``usual'' sensitivity to initial conditions can provide a more fundamental limit on predictability than expected, not just beyond some future but  at any time.

Let us now construct a numerical evidence.
Figure~\ref{Fig3} shows a ``mosaic''  representation of a subset of
sequences generated from Eq.~(\ref{Eq7}), with  $p=3$, $q=2$ and $m=9$,  $s=8$, for different values of $\theta$ as parameterized by  $k$ 
(recall that  $\theta$ encodes the initial condition for a given realization of the
series). We can observe that  there exists a  number of sequences $\{x^{(k,9,8)}_n\}_{n=8}^{17}$ that {\it coalesce} to the same values independently of $k$.
This is the ``band'' formed between $n=s=8$ and $n=s+m=17$. For a fixed time-step $n> m+s$ (i.e., forward in time) there exist horizontal (i.e., at $n$ fixed)
periodic structures whose periods are $T_{f}=q^{n-s-m}$. For example,  $x^{(k,9,8)}_n$ has two possible values when $n=18$,  four when $n=19$, eight when $n=20$, and so on. This is so because
the periods are $T_f=2,4,8,\ldots, 2^{n-s-m}$, respectively. In our present analysis, using the example of Fig.~\ref{Fig3} we have assessed 
that even though $x_n^{(k,9,8)}$
have the same values for every $0\leq k\leq100$
at any time between $8\leq n\leq 17$, 
there exist two possible values for the next time step $\{x_{18}^{(k,9,8)}\}_{k\in[0,100]}$.  
On the other hand, for a fixed time-step $n< s$ (i.e., backward in time) there also exist horizontal 
periodic structures but now the periods are $T_{b}=p^{s-n}$. From Fig.~\ref{Fig3} 
it is easy to verify that $T_b=3,9,27,\ldots, 3^{s-n}$.
The increasing horizontal periods  guarantee that each realization
of the sequences  (parameterized by $k$) is unique and independent from each other.
In this example, once we arrive at $n=8$, it is not possible to reconstruct the past of the sequence like 
an apparent {\it loss of memory}.
In Fig.~\ref{Fig3}, we also illustrate the branching 
forward in time. Starting from a unique value at $x^{(k,9,8)}_{17}$, the successive time steps ``bifurcate'' in a tree ramified 
between two possible values. 
As the horizontal period increases for successive time steps, it is possible to generate  
endless different sequences from Eq.~(\ref{Eq6aa}), even if they start  with the ``same initial 
condition''. 
Let us remark that these observations are also explained   in
terms of the images and pre--images under $\mathcal{F}_{3/2}$ in Eq.~(\ref{Eq6aa}), i.e., for a given value of $x$ there
are $q$--possible images and $p$--possible pre--images (see Figs.~\ref{Fig2} and \ref{Fig3}).
Moreover, if we apply the same analysis for $\beta\in\mathbb{N}_{>1}$, all the vertical sequences are the same, 
and we conclude the conceptual evidence of our argument.

\section{Chebyshev dynamical systems with $\beta$ irrational}\label{irrational}
\subsection{Algebraic irrational values}
\begin{figure}[!ht]
\begin{center}
\includegraphics[angle=0,width=0.5\linewidth]{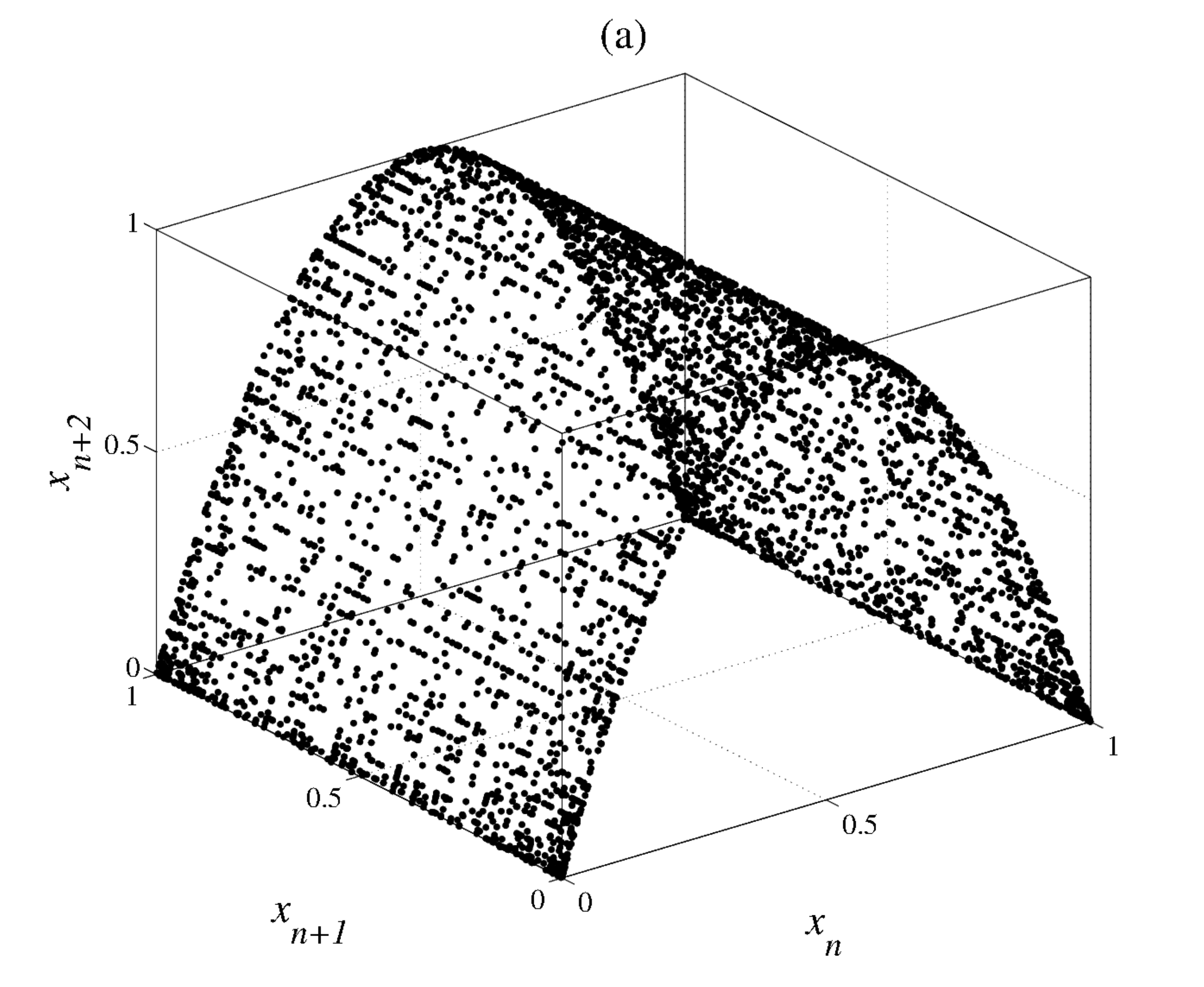}
\includegraphics[angle=0,width=0.5\linewidth]{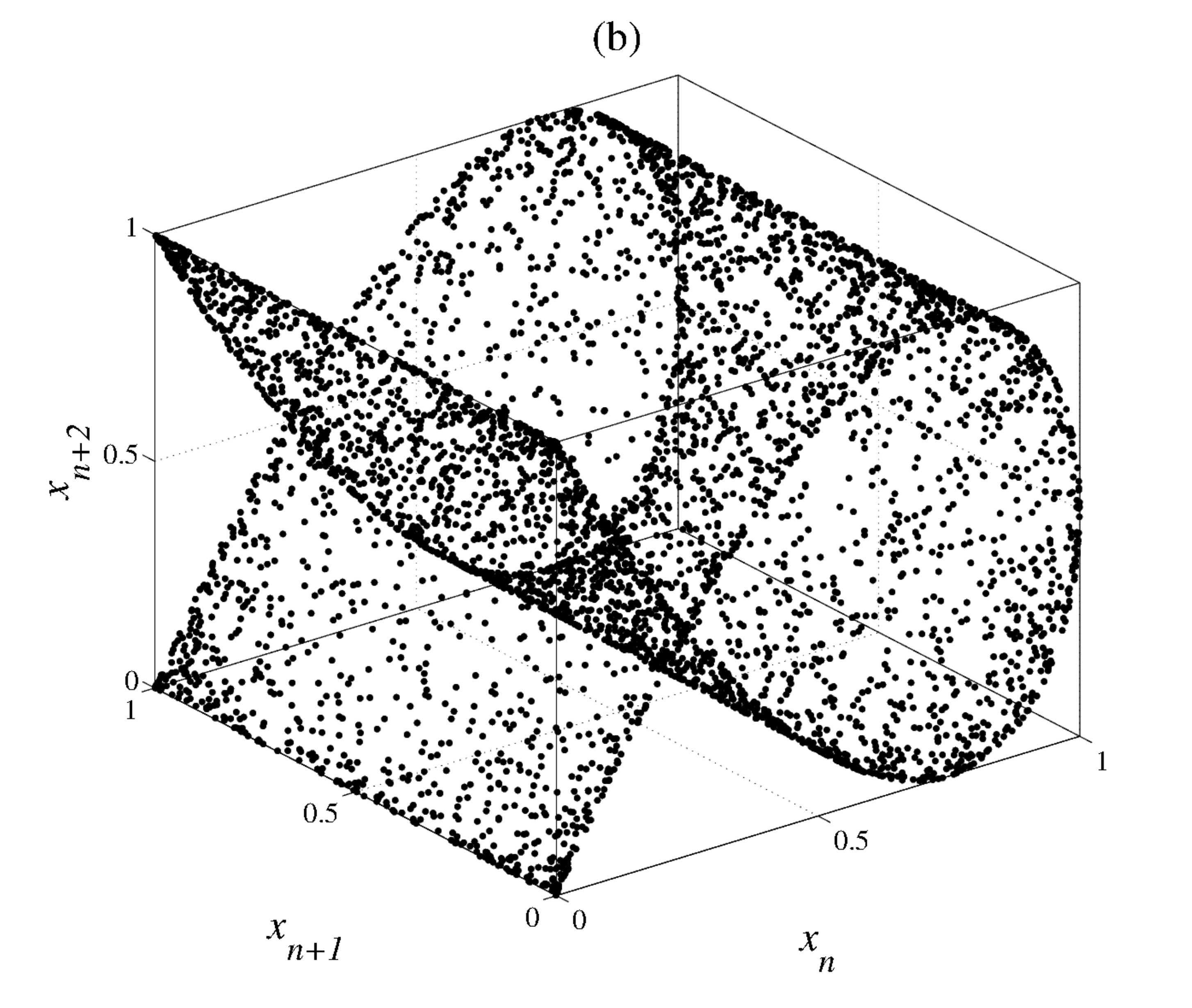}
 \caption{\label{FigSQRT} 
Three dimensional inverse cobweb plots for two examples of algebraic irrational values of the parameter $\beta\in(1,2)$, $\theta=\pi$ and (a) $\beta=\sqrt{2}$; (b) $\beta=\sqrt{3/2}$.
Calculated with Eq.~(\ref{Eq4}) and an arbitrary-precision arithmetic of $4\times 10^3$ digits for $\beta$.}
\end{center}
\end{figure}

In the same spirit of Section \ref{UvN} we use graphical displays to uncover the behavior of Eq.~(\ref{Eq4}) when $\beta$ is an algebraic irrational.
If we represent a two dimensional cobweb plot $(x_{n},x_{n+1})$  we  see a ``uniformly'' spread cloud of points (alike Fig.~\ref{Fig3new} (d) to be explained latter).
However, if we represent a multi-dimensional cobweb plot  as in Fig.~\ref{FigSQRT}, we can see  some known structures (the proper dimension depends on $\beta$).
For instance, with $\beta=\sqrt{2}$, we  see the graph of the logistic map ($\beta=2$) provided we look at the pairs $(x_n,x_{n+2})$.
Similarly  with $\beta=\sqrt{3/2}$, we  see the graph of the Fig~\ref{Fig1}(a).

Indeed, from a careful analysis of the dynamics from Eq.~(\ref{Eq6aa}), we can prove that for any real $\beta>1$ it holds that 
\begin{equation}
x_{n+k}=\sin^2(\beta^k\arcsin\sqrt{x_n} +\pi(\beta^k\omega_n - \omega_{n+k}))\,\, \forall n, k\geq 0.
\end{equation}
Therefore, if $\beta=\alpha^{1/k}$ for some integer $k\geq 1$ and rational $\alpha\in\mathbb{Q}_{>1}$, then
\begin{equation}
x_{n+k}=\sin^2(\alpha\arcsin\sqrt{x_n} +\pi\alpha\omega_n)\,\, \forall n\geq 0.
\end{equation}
In general, for $\alpha=p/q$, with $1\leq q<p$ coprime,   the graph corresponding to $\{(x_n,x_{n+k})\}_{n \in \mathbb{N}}$ is the graph of a Chebyshev dynamical system.
In particular, if $\alpha$ is an integer we then obtain  the  graph of a Chebyshev polynomial.

\begin{figure}[ht!]
\begin{center}
\includegraphics[angle=0,width=1.1\linewidth]{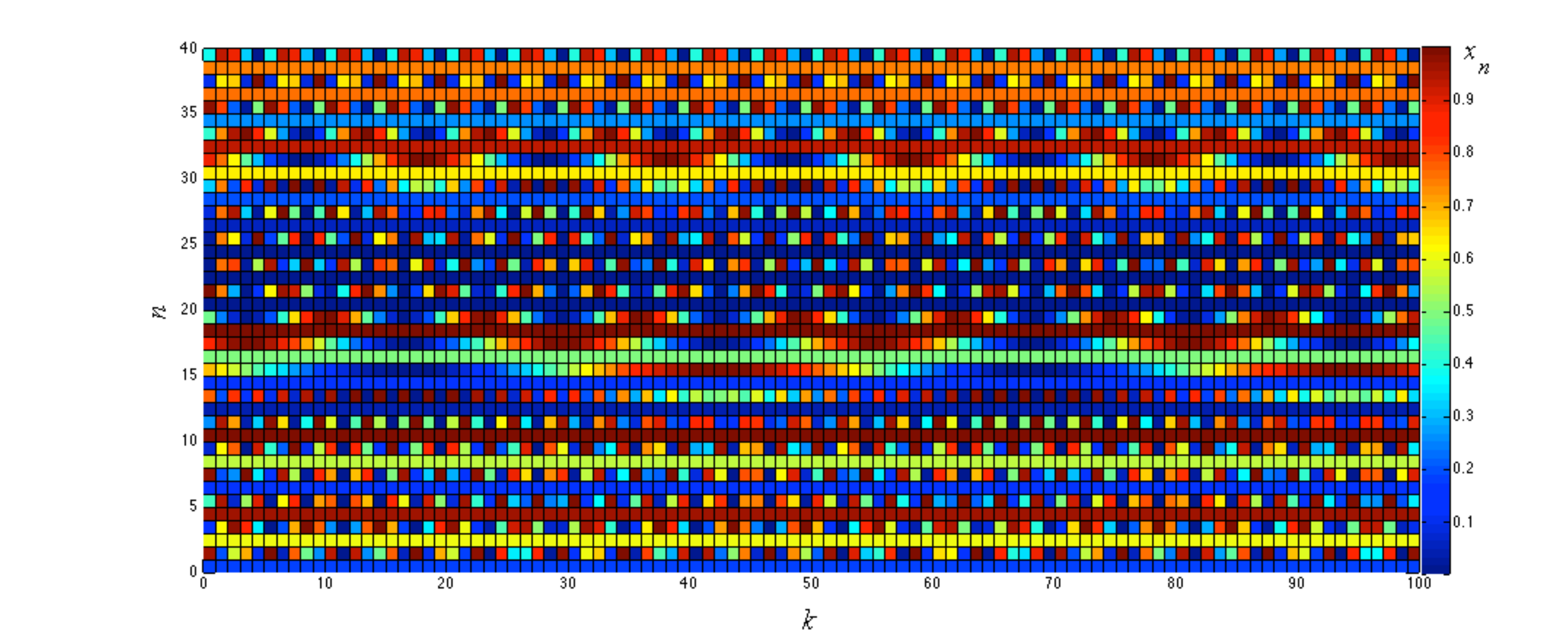}
 \caption{\label{mosaicoIrr} 
 Coalescence and branching of values  
$x_n^{(k,m,s)}$ after Eq.~(\ref{Eq7}) 
for $\beta=\sqrt{2}$, with $k\in[0,100]$, $m=0$, $s=0$, and $\theta_0=\pi$. The color of each tile in the mosaic
represents the value of $x^{(k,0,0)}_n\in[0,1]$.
}
\end{center}
\end{figure}

We have already mentioned in Subsection \ref{Coala} the possibility to have coalescence windows of arbitrary size if we take $\beta$ rational bounded and $q \to \infty$.
We also stressed that in such case a coalescence window can happen at most once.
However, as shown in Fig.~\ref{mosaicoIrr}, when $\beta$ is irrational, there exists the possibility of having 
coalescence windows of size one
distributed over $\mathbb{N}$
with a certain period.
With $\beta = \alpha^{1/k}$ for some integers $k\geq1$ and $\alpha >1$, the period is precisely $k$.

\subsection{Transcendental irrational values}
\begin{figure}[ht!]
\begin{center}
\includegraphics[angle=0,width=0.5\linewidth]{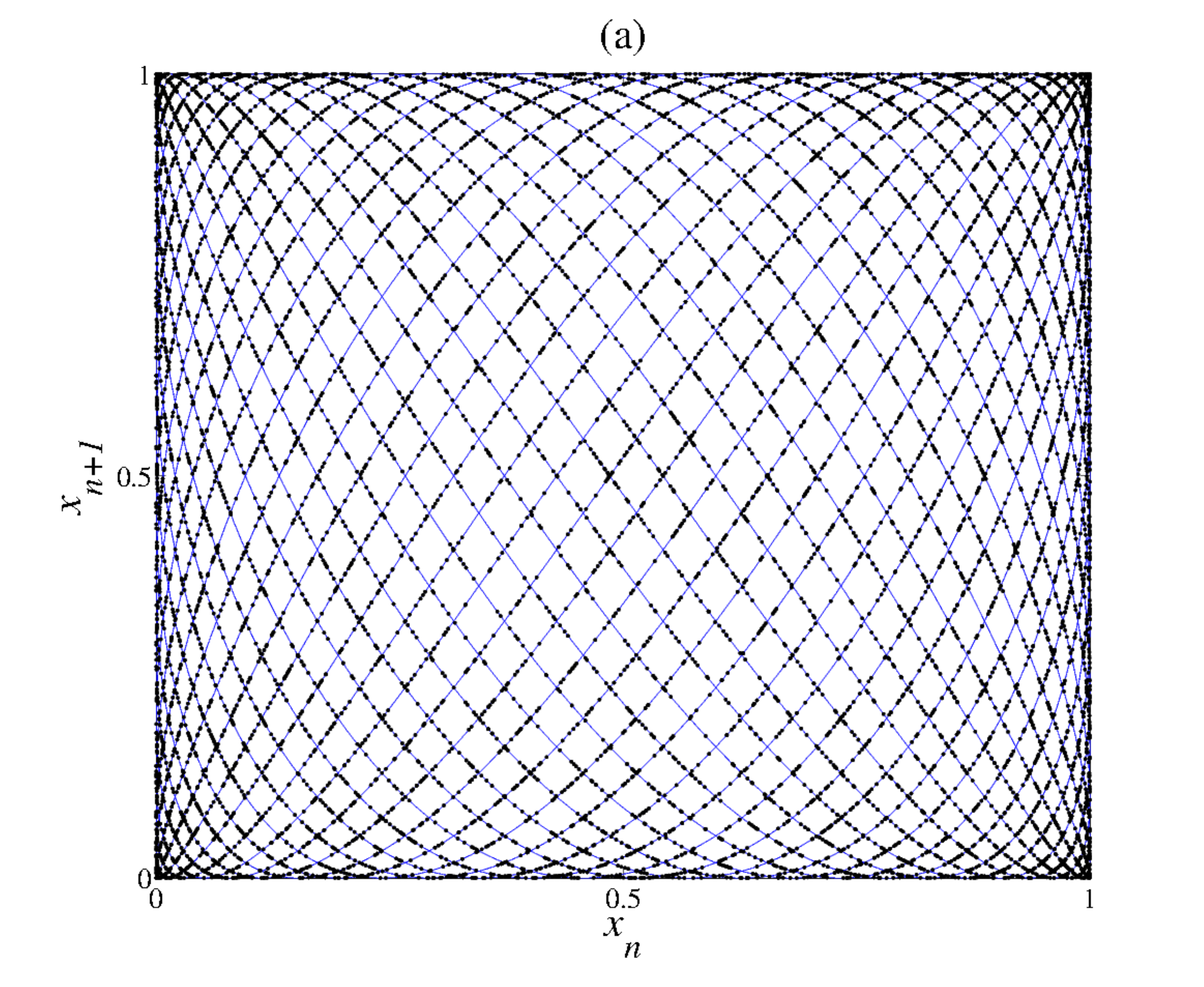}
\includegraphics[angle=0,width=0.5\linewidth]{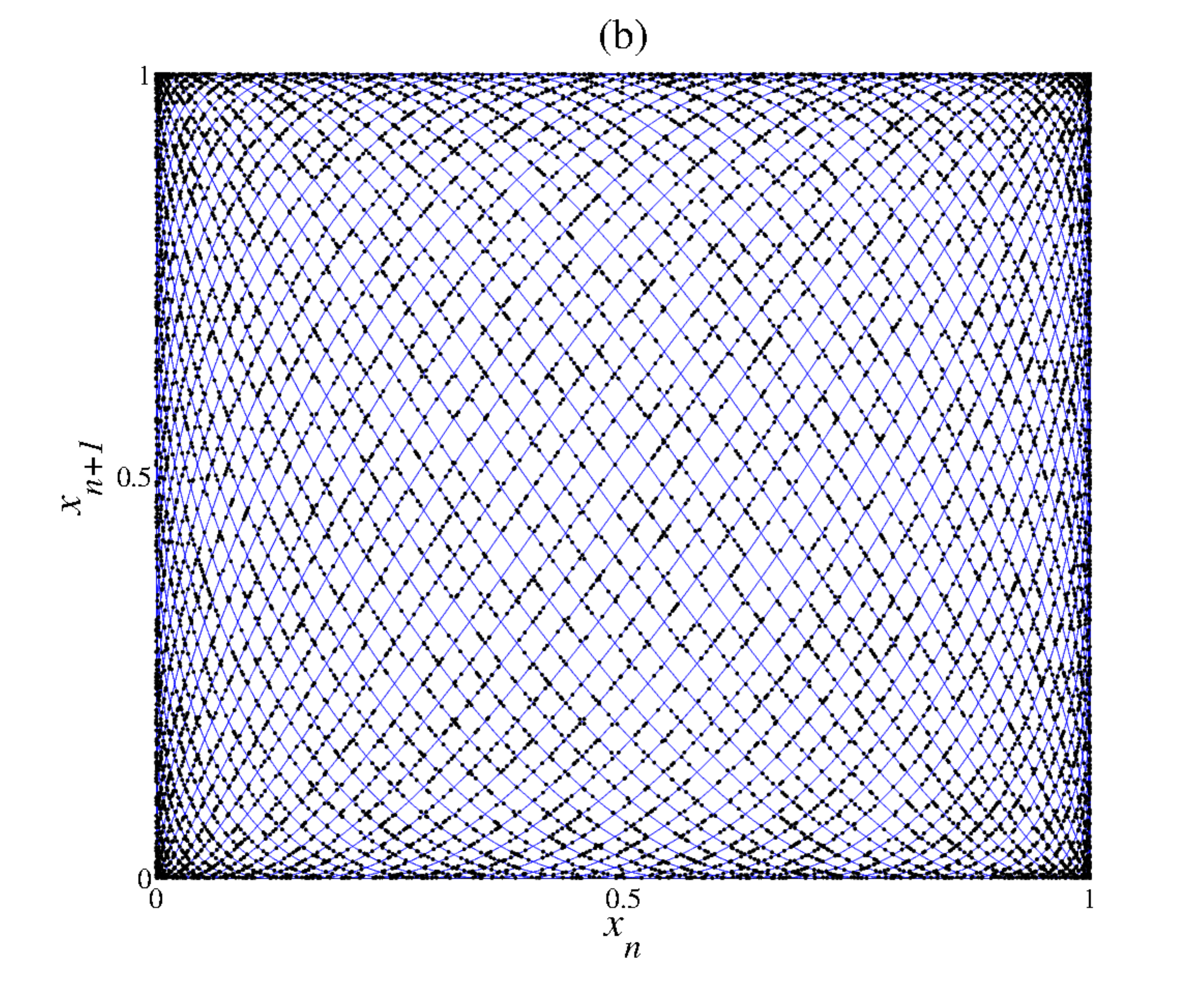}\\
\includegraphics[angle=0,width=0.5\linewidth]{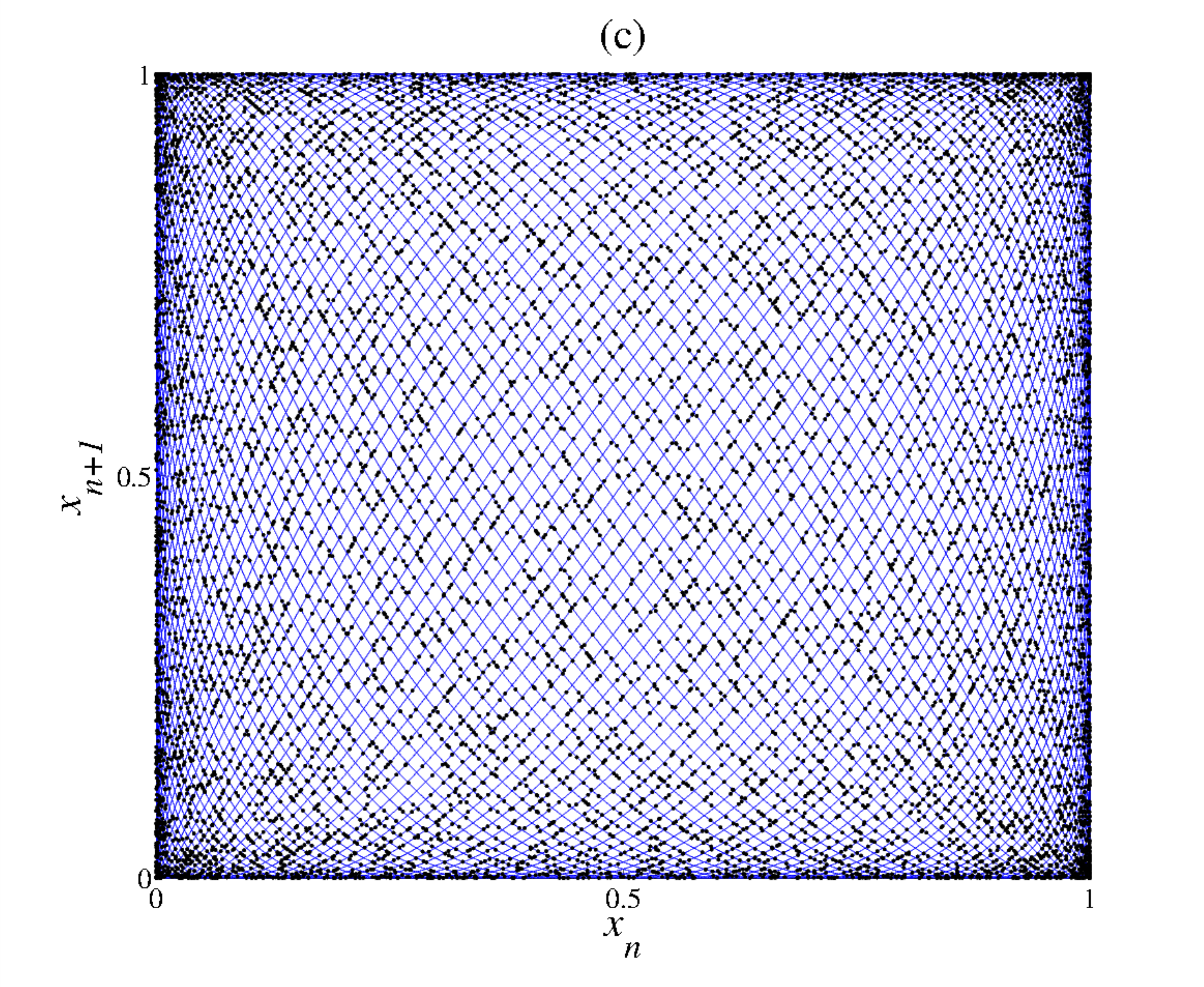}
\includegraphics[angle=0,width=0.5\linewidth]{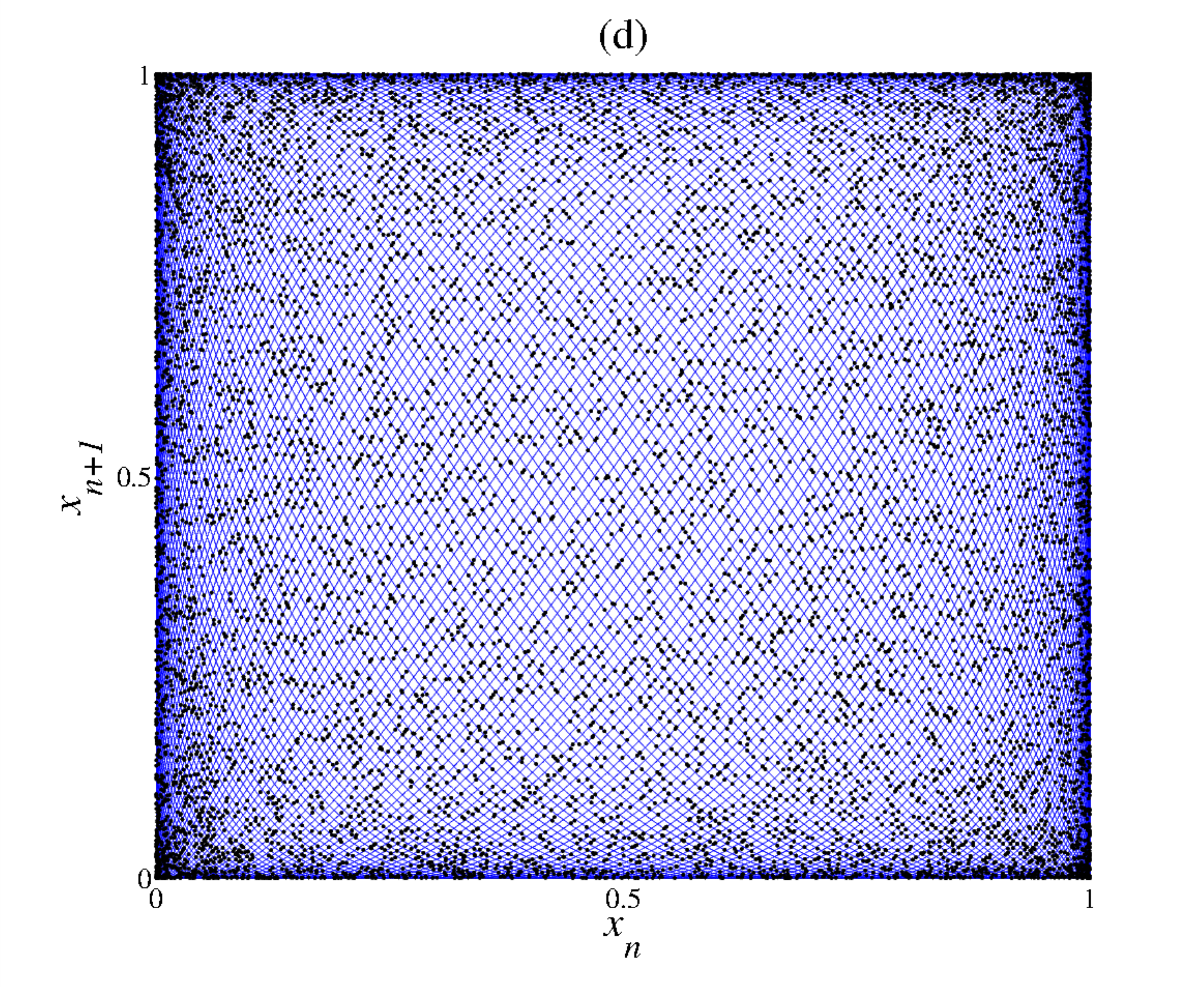}
 \caption{\label{Fig3new} 
 Four  successive approximations of the golden mean  calculated with Eq.~(\ref{golden}) as 
(a) $\beta_9=55/34;$ (b) $\beta_{10}=89/55/3$; (c) $\beta_{11}=144/89$; (d) $\beta_{12}=233/144$. The dots correspond to calculations made using Eq.~(\ref{Eq4}) and
the continuous blue lines with the corresponding function system (\ref{Eq6aa}).
}
\end{center}
\end{figure}

Now let us explore the case when $\beta$ is a transcendental irrational number.
The observation made in the previous subsection cannot happen in such case, i.e., there is no multidimensional inverse cobweb plot that can exhibit some regular graph (as in Fig~\ref{FigSQRT}).

However, we can infer some  properties when $\beta$ is
approximated as a continued fraction expansion, hence with a rational number. 
This approach permits us to control the representation of irrational numbers
like $\pi, e,\log 2$, etc,  up to some desired precision. 
We know from  number theory  that
every infinite simple continued fraction originated from an irrational number converges to that number.

As before, we use graphical displays to explore the case when $\beta$ is the golden mean with continued fraction expansion:
\begin{equation}\label{golden}
\beta = \frac{1+\sqrt{5}}{2}=1 + \cfrac{1}{1+\cfrac{1}{1 + \cfrac{1}{1 +  \cdots }}}.
\end{equation}

In Fig.~\ref{Fig3new}, we plot  the function system   $\mathcal{F}_\beta$ given by (\ref{Eq6aa})  for
four successive approximations (from ninth to twelfth convergent) of the golden mean $\beta_9 = 55/34$ (a); $\beta_{10}=89/55$ (b); $\beta_{11}=144/89$ (c) and $\beta_{12}=233/144$ (d) calculated from the 
expansion (\ref{golden}) (continuous blue lines). 
In the same Fig.~\ref{Fig3new} we also plot the sequences of $10^4$ points using Eq.~(\ref{Eq4}) and the same four successive approximations of the golden mean (black dots).

We can appreciate that the patterns  become 
more complicated as we take increasing convergents, with the blue curves filling more densely the square and the black dots distributing uniformly on the curves.
We can infer that to the limit, when $\beta$ is an irrational number, we have a system with infinitely many functions.
We can actually prove that the same mathematical framework of function systems includes and allows us to describe the dynamics for any real value of $\beta >1$~\cite{MT17}.

\section{Concluding remarks}\label{Conc}

The statements presented in this paper provide  numerical evidence of chaotic dynamical systems with
an 
uncertainty that goes beyond the uncertainty originating from the sensitivity to initial conditions in the sense of Lyapunov instability alone. 
We have pursued this through a consistent generalization of the seminal ideas developed by Kac, Ulam and von Neumann by extending the class of Chebyshev polynomial maps for $\beta \in \mathbb{N}_{>1}$ to a more general class of systems with real $\beta > 1$. 
Our findings show a hidden mathematical structure, which in a consistent formalism generalizes those canonical examples. 
We have demonstrated that Eq.~(\ref{Eq4}) corresponds to the closed-form expression of the solutions of pseudo-one-dimensional, still autonomous, recurrence relations that use function systems given by Eq.~(\ref{Eq6aa}).
%
We have also evidenced that the difficulty in the analysis of the original problem given in terms of real valued sequences $\{x_n\}_{n \in \mathbb{N}}$ is 
related to another known problem on discrete valued sequences $\{\omega_n\}_{n \in \mathbb{N}}$~\cite{KN75}.
Indeed, a straightforward application of Eqs. (\ref{Eq4}) and (\ref{Eq6aa}) gives a new and unexpectedly simple construction of deterministic function systems~\cite{A98,F03}.
They are defined as the selection at every $n \geq 0$ of a mapping $f_{\omega_n}$ by which a given value $x_ n$ is updated to $x_{n+1} = f_{\omega_n}(x_n )$, $\omega_n \in \Omega_\beta \subset \mathbb{N}$, through a deterministic selection mechanism in a non-hyperbolic family of nonlinear, possibly infinite, function systems.
This finding closes the arguments presented in Ref.~\cite{M04} against the lack of predictability in the sequences (\ref{Eq4}) due to the (apparent) nonexistence of a nonlinear recursion relation.

We have shown that the Chebyshev dynamical systems produce arbitrarily long and complex sequences, and could
provide some insights into how to build a deterministic process that behaves more closely to a truly random dynamics. 
The very physical origin of (and/or the relevance of what would be a mathematical artifact for) such models, deserves more considerations, in particular as hidden variable models.
Notwithstanding, we can sketch some examples where the insights addressed in this work could already be useful for other disciplines. 
First, it is interesting to note that (iterated) function systems can be used to study
%
many real life processes like digital communications channels~\cite{BHMS04}, history-dependent neural representation at the Hippocampus~\cite{KFYTT09}, turbulent fluid flows~\cite{HORS16}, actuarial science~\cite{CT08}, and even nonunitary quantum dynamics~\cite{LZS03}. 
Also, from the point of view of numerical simulations, it should be worth to explore the use of function systems such as Eq.~(\ref{Eq6aa}) as an alternative for pseudo-random numbers generator algorithms.
It may also be of interest for the design of pseudo-random functions primitives and to enforce security  in chaos-based cryptography~\cite{K11}.
Finally, in cognitive and behavioral sciences the notion of randomness seems to be crucial to emulate free behavior in embodied systems~\cite{B12}. 
In all these frameworks the issue of model misspecification arises, i.e., from time series observations, one can easily come to formulate either a probabilistic model for the choice among the branches of the function system, or decide to consider a multivalued map model~\cite{DF99}. 
In either case, such formulation would provide a valuable statistical description of the dynamics on ensembles of states, but also very likely introduce spurious orbital behaviors, and would inevitably and irreversibly conceal the self-determinism in the selection process. 
After introducing such a bias, there is no objective criteria left to decide what is subjected to chance and what is to self-determinism.

\section*{Acknowledgments}
We thank H.P. de Vladar for discussions that have contributed to the insights presented here.
The assistance of A. Maneiro is friendly recognized. 
L.T. is grateful to J.A. Gonz\'alez for the invaluable past collaboration on this subject.
A.M. is grateful to P. Guiraud, E. Ugalde and M.I. Cortez for enlightening discussions.


\end{document}